\documentclass[preprint,prb,showpacs,amsmath,amssymb]{revtex4}

\usepackage[dvips]{graphicx}
\usepackage{pstricks}
\usepackage{dcolumn}
\usepackage{bm}

\newcommand{\imag}{{\rm i}}

\newcommand{\sgn}{{\rm sgn}}

\begin{document}

\title{Strong-coupling approach
to the Mott--Hubbard insulator \\
on a Bethe lattice in Dynamical Mean-Field Theory}

\author{Daniel Ruhl}
\author{Florian Gebhard}
\affiliation{Department of Physics, 
Philipps-Universit\"at D-35032 Marburg, Germany}

\date{\today}

\begin{abstract}
We calculate the Hubbard bands for the half-filled Hubbard model
on a Bethe lattice with infinite coordination number up to
and including third order in the inverse Hubbard interaction.
We employ the Kato--Takahashi perturbation theory 
to solve the self-consistency equation of the Dynamical Mean-Field Theory 
analytically for the single-impurity Anderson model in
multi-chain geometry. The weight of the secondary Hubbard sub-bands
is of fourth order so that the two-chain geometry is sufficient for our study.
Even close to the Mott--Hubbard transition,
our results for the Mott--Hubbard gap agree very well with those
from numerical Dynamical Density-Matrix Renormalization Group (DDMRG)
calculations. The density of states of the lower Hubbard band
also agrees very well with DDMRG data, apart from a resonance contribution
at the upper band edge which cannot be reproduced in low-order 
perturbation theory. 
\end{abstract}


\maketitle

\section{Introduction}
\label{sec:intro}

The Dynamical Mean-Field Theory (DMFT) maps lattice models 
for electrons with a Hubbard-type interaction 
onto effective single-impurity models;
for a review, see Ref.~[\onlinecite{dmftreviews}]. 
The parameters of the impurity model must be determined in such a way
that the self-energy and the Green function of the impurity model 
agree with the local self-energy and the local Green function 
of the lattice model. The solution of this 
self-consistency problem equally solves the original lattice problem 
in the limit of infinite dimensions.~\cite{MV}
For example, the Hubbard model on the Bethe lattice with infinite
coordination number can be mapped
onto the single-impurity Anderson model. Then, the self-consistency  
condition requires that its hybridization function and its
Green function agree for all frequencies.

Unfortunately, we are far from an analytical solution of 
the single-impurity Anderson model for a general hybridization function,
and a variety of methods have been employed to solve
the DMFT equations for the single-band Hubbard model. Examples
for numerical treatments are
the Numerical Renormalization Group method,~\cite{NRG}
Exact Diagonalization,~\cite{ED1,ED2,kalinowski} the
Random Dispersion Approximation,~\cite{kalinowski,RDA}
the Dynamical Density-Matrix Renormalization Group (DDMRG) 
method,~\cite{kalinowski,Nishimoto,Uhrig} and, at finite temperatures,
Quantum Monte-Carlo.~\cite{Jarrell,Bluemer,Rubtsov}
Approximate analytical methods at zero temperature 
include the Iterated Perturbation 
Theory~\cite{IPT}, the Local Moment Approach,~\cite{LMA} and
the self-energy functional approach.~\cite{Potthoff}

All methods have their merits and limitations and it is desirable
to compare their results with those from perturbation theory.
For the half-filled Hubbard model on a Bethe lattice
with infinite coordination number,
the self-energy~\cite{Mahlert} and the ground-state 
energy~\cite{danieldiplom} are known up to and including
fourth order in the Hubbard interaction~$U$
and up to second order in $1/U$.~\cite{kalinowski}
However, these calculations are based on the Hubbard model
in infinite dimensions, not on the DMFT description.

In this work, we solve the DMFT equations for the Hubbard model 
on a Bethe lattice with infinite coordination number, $Z\to\infty$,
at half band-filling for strong coupling
where the model describes a paramagnetic Mott--Hubbard insulator.
Up to and including third order in $1/U$, 
we determine the hybridization function 
of the single-impurity Anderson model which corresponds to
the Hubbard model on the $Z\to\infty$ Bethe lattice. 
Essential to our approach are:
(i)~the mapping of the single-impurity Anderson model from the `star geometry'
onto the `multi-chain geometry' where each chain represents 
one of the upper and lower Hubbard sub-bands;
(ii)~the Kato--Takahashi perturbation theory~\cite{kato,takahashi}
for degenerate ground states; (iii)~the Lanczos representation
of the hybridization function and the Green function which permits
an order-by-order solution of the self-consistency equation
for the moments of the density of states;
(iv)~the locality of the Hubbard interaction and of the Lanczos operators
in finite order perturbation theory.

Our work is organized as follows. In Sect.~\ref{sec:model}
we introduce the Hubbard model, the single-impurity Anderson model,
the DMFT equations which link the two models, and the two-chain mapping
which we use for our perturbative calculations to third order in~$1/U$.
In Sect.~\ref{sec:KT-method} we adapt the Kato--Takahashi 
perturbation theory to our problem and use the Lanczos algorithm
to express the density of states for the (primary) lower Hubbard band 
and for the hybridization function in terms of their moments.
Then, the self-consistency equation reduces to the condition that
the respective moments agree up to trivial signs.
In Sect.~\ref{sec:results} we investigate the lowest non-trivial
order and show how the iterative solution of the DMFT equation works in practice.
Next, we summarize the results to third order; all technical details
can be found in Ref.~[\onlinecite{ruhl}].
The remaining problem is the calculation of the density of states
at the boundary of a semi-infinite chain for a single particle
which can move between nearest neighbors and experiences a 
local potential at and near the boundary.
Its solution and a favorable 
comparison with
previous numerical work~\cite{Nishimoto,Bluemer} is the subject of
Sect.~\ref{sect:thirdorderbands}.
Conclusions, Sect.~\ref{sec:conclusions}, and two appendices,
on the secondary Hubbard bands and on the Green functions for
a particle on a semi-infinite chain, close our presentation.

\section{Mott--Hubbard insulator in Dynamical Mean-Field Theory}
\label{sec:model}

We start our presentation with the definition of 
the Hubbard model and the single-impurity Anderson model.
For a specific choice of the hybridization function in the
single-impurity Anderson model, 
its single-particle Green function 
is identical to the local single-particle Green function
of the Hubbard model in infinite dimensions.
The Dynamical Mean-Field Theory prescribes a way
to determine the hybridization function self-consistently.

In general, the single-particle Green function for the single-impurity Anderson
model cannot be calculated analytically.
For the Mott--Hubbard insulator, 
we use a mapping of the
model onto a multi-chain geometry where 
the chains represent the energy levels 
in the energetically separated upper and lower Hubbard sub-bands.

\subsection{Hamilton operators and Green functions}

\subsubsection{Hubbard model}
We consider the repulsive single-band Hubbard model ($U\geq 0$)
\begin{equation}\label{eq:hubbardhamiltonianplain}
\hat{H}  = \sum_{i,j;\sigma}  t_{ij} \hat{c}_{i, \sigma}^+ \hat{c}_{j, \sigma} 
+ U \sum_{i} \hat{n}_{i, \uparrow} \hat{n}_{i, \downarrow} 
-\mu \sum_{i}\left(\hat{n}_{i, \uparrow} + \hat{n}_{i, \downarrow}\right) 
=: \hat{T} + U \hat{D} -\mu \hat{N}\; .
\end{equation}
Here, $\hat{T}$ denotes the operator for the electron transfer between
the lattice sites $i$ and~$j$,
the fermion operator $\hat{c}_{i, \sigma}^+$ ($ \hat{c}_{i, \sigma}$) creates
(annihilates) an electron with spin~$\sigma$ ($=\uparrow,\downarrow$)
on lattice site~$i$,
the operator $\hat{n}_{i,\sigma}=\hat{c}_{i, \sigma}^+\hat{c}_{i, \sigma}$
counts the number of $\sigma$-electrons on site~$i$, and
the operator 
$\hat{D}= \sum_{i} \hat{n}_{i, \uparrow} \hat{n}_{i, \downarrow}$
counts the number of doubly occupied sites. 
For the description of the Mott--Hubbard insulator, 
we consider a half-filled system 
where there is on average one electron per lattice site, $n=N/L=1$.
Moreover, we treat the paramagnetic situation, $n_{\uparrow}=n_{\downarrow}=1/2$,
without any symmetry breaking. 
The thermodynamic limit, $N,L\to\infty$ 
is implicit in our calculations below. 

As a major simplification, we assume that the electrons move
between nearest neighbors on a Bethe lattice with coordination number $Z$,
\begin{equation}
t_{ij}= \left\{ \begin{array}{cll}
-t/\sqrt{Z} & \hbox{if} & \hbox{$i$, $j$ are nearest neighbors}\; , \\
0 & \hbox{else}\; .
\end{array}
\right. 
\end{equation} 
Later, we shall let go $Z\to\infty$ and choose $t=1$ as our unit of energy.
The Bethe lattice with coordination number $Z$
is an infinite $Z$-Cayley tree. 
A $Z$-Cayley tree is constructed
from a first site by connecting it to $Z$~new sites which constitute
the first shell. One creates further shells by adding $Z-1$ new sites 
to every site in shell~$s$. The Cayley tree has no loops and all closed paths
are self-retracing.~\cite{baxter}
The Bethe lattice contains $s\to\infty$ shells.
Since the Bethe lattice is bipartite, 
the chemical potential $\mu=U/2$ guarantees
half band-filling at all temperatures.

We are interested in the local Green function of the Hubbard model
in its exact ground state $|\Psi_0\rangle$. We use the abbreviation
\begin{equation}
\langle \hat{A}\rangle =
\frac{\langle \Psi_0 | \hat{A}| \Psi_0\rangle}{\langle \Psi_0 | \Psi_0\rangle}
\end{equation}
for ground-state expectation values and define
the local causal Green function 
in the time domain
\begin{equation}
G_{\sigma}(i;t) = -\imag \langle \hat{T}_{\rm s} 
\hat{c}_{i, \sigma}(t)\hat{c}_{i, \sigma}^+(0)\rangle \;,
\end{equation}
where the Heisenberg operators
\begin{equation}
\hat{c}_{i,\sigma}(t)=e^{\imag \hat{H} t}\hat{c}_{i,\sigma}e^{-\imag \hat{H} t}
\end{equation}
are time-ordered with the help of time-ordering operator $\hat{T}_{\rm t}$
\begin{equation}
	\hat{T}_{\rm t} \hat{c}_{i,\sigma}(t) \hat{c}_{j,\sigma'}^+(t') 
= \left\{ \begin{array}{rcl}
\hat{c}_{i,\sigma}(t) \hat{c}_{j,\sigma'}^+(t') & \hbox{for} & t>t' \; ,\\
- \hat{c}_{j,\sigma'}^+(t') \hat{c}_{i,\sigma}(t) & \hbox{for} & t<t' \; .
\end{array}
\right.
\end{equation}
The time-frequency Fourier transformation 
of the local Green function is defined as
\begin{eqnarray}
G_{i,\sigma}(\omega) &=& 
\int_{-\infty}^{\infty} {\rm d} t\, e^{\imag \omega t} G_{\sigma}(i;t) \\
\label{eq:fouriertransformlocalgreen}
& =&  \lim_{\eta \to 0^+} \Big \{ 
\langle \hat{c}_{i,\sigma} 
\left(\omega - (\hat{H} - E_0(N)) + \imag \eta\right)^{-1} 
\hat{c}_{i,\sigma}^+\rangle  
 \nonumber \\
 && 
\hphantom{\lim_{\eta \to 0^+} \Big \{ }
+ \langle \hat{c}_{i,\sigma}^+ 
\left(\omega +(\hat{H} - E_0(N))-\imag \eta\right)^{-1} \hat{c}_{i,\sigma}
\rangle \Big\} \; .
\label{etaforthefirsttime}
\end{eqnarray}
The limit $\eta\to 0^+$ is implicitly understood henceforth.
In~(\ref{etaforthefirsttime}), $E_0(N)$ denotes the energy of the 
$N$-particle ground state $|\Psi_0\rangle$ of the Hubbard model.
The (local) density of states is obtained from the imaginary part of the
Green function ($\sgn(x)$ is the sign function),
\begin{equation}
D_{\sigma}(\omega) = 
-\frac{1}{\pi} \sgn(\omega) {\rm Im} \left[G_{i,\sigma}(\omega)\right] \; .
\end{equation}
The density of states is positive semi-definite 
and its integral over all frequencies
is unity.~\cite{FetterWalecka}

The Green function for non-interacting electrons on a Bethe lattice
($U=0$) can be calculated in various ways.~\cite{Kollar}
In the limit $Z\to\infty$, it approaches the Hubbard semi-ellipse,
\begin{equation}
\rho(\omega) 
= \frac{4}{\pi W} \sqrt{1- \Big ( \frac{2\omega}{W} \Big )^2}
\quad \hbox{for} \quad |\omega|\leq W/2 
\label{eq:semiellipse}
\end{equation}
with $W=4$ as the bare bandwidth. 
In the presence of interactions ($U>0$), 
the local Green function can be expressed
with the help of the (proper) self-energy
$\Sigma_{\sigma}(\omega)$,
\begin{equation}
\label{HubbardGreenlocal}
G_{i,\sigma}(\omega)= \int_{-\infty}^{\infty} {\rm d} \omega'
\frac{\rho(\omega')}{\omega-\omega'-\Sigma_{\sigma}(\omega)}=
G_{i,\sigma}^{(0)}(\omega-\Sigma_{\sigma}(\omega))\; .
\end{equation}
Note that, in the limit $Z\to\infty$, the self-energy depends only 
on the frequency.~\cite{MV} In principle, the self-energy can be calculated in 
diagrammatic perturbation theory.~\cite{FetterWalecka}

\subsubsection{Single-impurity Anderson model}

In order to set up the Dynamical Mean-Field Theory for the half-filled, 
paramagnetic Hubbard model,
we consider the discrete, symmetric single-impurity Anderson model
(SIAM) in `star geometry',
\begin{equation}
\label{eq:SIAM-H}
	\hat{H}_{\rm SIAM} = 
\sum_{m=0}^{L-2}\sum_{\sigma} 
\xi_m \hat{a}_{m,\sigma}^+ \hat{a}_{m,\sigma} 
-\frac{U}{2} \sum_{\sigma} \hat{n}_{d,\sigma} 
+ \sum_{m=0}^{L-2}\sum_{\sigma} V_{m} 
(\hat{a}_{m,\sigma}^+ \hat{d}_{\sigma} 
+ \hat{d}_{\sigma}^+ \hat{a}_{m,\sigma}) 	
+ U \hat{n}_{d,\uparrow} \hat{n}_{d,\downarrow} \; .
\end{equation}
Here, $\hat{a}_{m,\sigma}^+$ ($\hat{a}_{m,\sigma}$) 
creates (annihilates)
a bath electron with spin $\sigma$ and bath energy~$\xi_m$,
$\hat{d}_{\sigma}^+$ ($\hat{d}_{\sigma}$) 
creates (annihilates) an electron with spin $\sigma$ on the impurity level
with energy $E_d=-U/2$, and
$\hat{n}_{d,\sigma}=\hat{d}_{\sigma}^+\hat{d}_{\sigma}$ counts the number
of $\sigma$-electrons on the impurity. The Hubbard interaction on the impurity
is the same as in the Hubbard model. The parameters $V_{m}>0$ describe
the hybridization between the bath levels and the impurity site.
The half-filled case corresponds to $N=L$ electrons.
The limits $N,L\to\infty$ are implicitly understood henceforth.

The single-impurity Anderson model is fully characterized by
the hybridization function,~\cite{Hewson}
\begin{equation}
\Delta(\omega) = \sum_{m} 
\frac{V_{m}^2}{\omega - \xi_{m} + \imag\, \sgn(\omega) \eta} \; .
\label{deltaofomega}
\end{equation}
The (causal) Green function of the impurity electrons is defined by
\begin{equation}
G^{\rm SIAM}_{\sigma}(d;t) = -\imag \langle \hat{T}_{\rm s} 
\hat{d}_{\sigma}(t)\hat{d}_{\sigma}^+(0)\rangle \;,
\end{equation}
where the expectation value is to be taken in the exact ground state
of the single-impurity Anderson model.
After Fourier transformation, the Green function can be expressed with the help
of the hybridization function and the (proper) self-energy 
in the form~\cite{Hewson}
\begin{equation}
G^{\rm SIAM}_{\sigma}(\omega) = 
\int_{-\infty}^{\infty} {\rm d} t\, e^{\imag \omega t} 
G^{\rm SIAM}_{\sigma}(d;t) =
\frac{1}{\omega-\Delta(\omega)-\Sigma_{\sigma}^{\rm SIAM}(\omega)} \; .
\label{eq:G-Delta-Sigma}
\end{equation}
As in the case of the Hubbard model, the self-energy of the single-impurity
model can be calculated in diagrammatic perturbation theory.

\subsubsection{DMFT equations on the Bethe lattice}

The skeleton diagrams for the single-impurity Anderson model
and the Hubbard model with infinite coordination number are identical;
for a review, see Ref.~[\onlinecite{dmftreviews}].
Therefore, if their local Green functions agree, 
\begin{equation}
\label{oneDMFT}
G_{i,\sigma}(\omega) =G^{\rm SIAM}_{\sigma}(\omega) \; ,
\end{equation}
their self-energies agree as well,
\begin{equation}
\label{twoDMFT}
\Sigma_{\sigma}(\omega) =\Sigma^{\rm SIAM}_{\sigma}(\omega) \; .
\end{equation}
The exact solution of the Hubbard model for infinite coordination
number reduces to the
calculation of the Green function of the single-impurity Anderson model
for a general hybridization function $\Delta(\omega)$. The 
DMFT self-consistency equations~(\ref{oneDMFT}) and~(\ref{twoDMFT})
single out the hybridization function 
which describes the Hubbard model in the limit
of infinite coordination number.

For the Hubbard model on the Bethe lattice, the semi-elliptic
bare density of states~(\ref{eq:semiellipse}) 
results in the following 
form of the local Green function~(\ref{HubbardGreenlocal})
\begin{equation}
G_{i,\sigma}(z)=\frac{1}{2} \left(z-\sqrt{z^2-4}\right) 
\quad , \quad z=\omega-\Sigma_{\sigma}(\omega)
\end{equation}
so that
\begin{equation}
\Sigma_{\sigma}(\omega)=\omega - G_{i,\sigma}(\omega) 
-\frac{1}{G_{i,\sigma}(\omega)}
\end{equation}
holds. Together with~(\ref{eq:G-Delta-Sigma}), 
the DMFT equations~(\ref{oneDMFT}) and~(\ref{twoDMFT})
reduce to the single condition
\begin{equation}
\label{what-we-need}
\Delta(\omega) = G^{\rm SIAM}_{\sigma}(\omega)
\end{equation}
on the hybridization function.
The remaining task is to calculate the Green function 
$G^{\rm SIAM}_{\sigma}(\omega)$ for the single-impurity
Anderson model for a general hybridization function $\Delta(\omega)$.
The equation~(\ref{what-we-need}) singles out the hybridization function 
which describes the Hubbard model 
on the Bethe lattice with infinite coordination number. From now 
on we shall exclusively investigate the single-impurity Anderson
model. Therefore, we drop the superscript `SIAM' on all quantities.

\subsection{Two-chain mapping for the Mott--Hubbard insulator}

\subsubsection{Hubbard bands and charge gap}

We are interested in the description of the Mott--Hubbard insulator
where the charge gap separates the upper and lower Hubbard bands.
Due to particle-hole and spin symmetry (see, for example, 
Refs.~[\onlinecite{kalinowski,David}]), it is sufficient to calculate
the Green function of the lower Hubbard band for a fixed spin, 
say, $\sigma=\uparrow$,
\begin{equation}
G_{\rm LHB}(\omega<0) = \langle \hat{d}_{\uparrow}^+
\left( \omega +\hat{H}-E_0-\imag \eta \right)^{-1}
\hat{d}_{\uparrow} \rangle \; .
\label{eq:whatweneedtocal}
\end{equation}
For positive frequencies we have $G_{\rm UHB}(\omega>0)=-
G_{\rm LHB}(-\omega)$. Moreover, for the density of states we have
$D_{\rm UHB}(\omega>0)=D_{\rm LHB}(-\omega)$, i.e., the density of states
is symmetric around $\omega=0$.

The upper edge of the lower Hubbard band is the chemical potential~$\mu^{-}<0$
for adding the $L$th electron. 
The minimal energy for adding another electron
to the system ($N=L+1$), the chemical potential~$\mu^{+}$, 
is given by
$\mu^{+}=-\mu^{-}$.~\cite{LiebWu} Therefore, the charge gap
obeys
\begin{equation}
\Delta_{\rm c}= \mu^{+}-\mu^{-}= 2 |\mu^{-}|\;,
\label{eq:chargegap}
\end{equation}
i.e., it can be obtained from the upper band edge
of the lower Hubbard band.

\subsubsection{Two-chain single-impurity Anderson model}

The self-consistency equation~(\ref{what-we-need}) demands that the
imaginary part of the hybridization function is identical to
the density of states. For the Hubbard model at strong coupling, $U\gg W$,
we know that the density of states is centered in the two Hubbard bands,
$|\omega\pm U/2|\leq {\cal O}(W/2)$.~\cite{Mottbook}
For discrete bath levels, the imaginary part of the hybridization
function $\Delta(\omega)$ in eq.~(\ref{deltaofomega}) 
consists of peaks at the bath energies $\xi_m$ with
weights~$V_{m}^2$. Therefore, the bath energies can be grouped
into those of the lower Hubbard band, 
$\xi_m=-{\cal O}(U/2)$, and those 
of the upper Hubbard band, 
$\xi_m={\cal O}(U/2)$. Consequently, we map the single-impurity
Anderson model in star geometry, eq.~(\ref{eq:SIAM-H}),
onto a two-chain geometry where the impurity site hybridizes
with two sites which represent the lower and upper 
Hubbard bands.~\cite{Nahrgang}
Note that, in numerical treatments of the single-impurity Anderson model,
the star geometry is usually mapped onto a single chain.~\cite{dmftreviews}
Apparently, the two-chain mapping is more adequate for the
Mott--Hubbard insulator; a similar idea was proposed earlier
in Refs.~[\onlinecite{ED2,two-chain-rucki}].
The concept is readily generalized to a multi-chain 
mapping where each region with a finite density of states is represented
by its own chain; see below.

\begin{figure}[ht]
	\begin{center}
		\scalebox{0.8} 
		{
		\begin{pspicture}(0,-2.7280803)(5.0,2.7280803)
	\psline[linewidth=0.02cm](0.21520574,-1.84)(3.8952057,1.82)
	\psdots[dotsize=0.18](3.8752058,1.8)
	\psdots[dotsize=0.18](0.17520574,-1.88)
	\psline[linewidth=0.02cm](3.7796948,-1.7754748)(0.090236485,1.8749907)
	\psdots[dotsize=0.18,dotangle=90.460495](0.11039658,1.855152)
	\psdots[dotsize=0.18,dotangle=90.460495](3.820015,-1.815152)
	\psline[linewidth=0.02cm](2.0264819,-2.5525966)(2.064317,2.6374485)
	\psdots[dotsize=0.18,dotangle=44.73844](2.064188,2.6091645)
	\psdots[dotsize=0.18,dotangle=44.73844](2.0262237,-2.6091645)
	\psline[linewidth=0.02cm](3.014097,-2.3616414)(1.0438904,2.440055)
	\psdots[dotsize=0.18,dotangle=67.46519](1.0546983,2.413917)
	\psdots[dotsize=0.18,dotangle=67.46519](3.0357132,-2.413917)
	\psline[linewidth=0.02cm](1.0523362,-2.3516612)(3.0708625,2.4299235)
	\psdots[dotsize=0.18,dotangle=22.26928](3.0599334,2.403836)
	\psdots[dotsize=0.18,dotangle=22.26928](1.030478,-2.403836)
	\psline[linewidth=0.02cm](1.548396,-2.503853)(2.5582993,2.5871286)
	\psdots[dotsize=0.18,dotangle=33.93596](2.5528712,2.5593703)
	\psdots[dotsize=0.18,dotangle=33.93596](1.5375402,-2.5593703)
	\psline[linewidth=0.02cm](0.5913671,-2.15192)(3.5065117,2.1422539)
	\psdots[dotsize=0.18,dotangle=10.98509](3.4906893,2.1188095)
	\psdots[dotsize=0.18,dotangle=10.98509](0.5597222,-2.1988094)
	\psline[linewidth=0.02cm](2.5099626,-2.5462143)(1.5241045,2.5494783)
	\psdots[dotsize=0.18,dotangle=56.105793](1.5295526,2.5217237)
	\psdots[dotsize=0.18,dotangle=56.105793](2.5208588,-2.6017237)
	\psline[linewidth=0.02cm](3.3969035,-2.1467838)(0.52638626,2.1773498)
	\psdots[dotsize=0.18,dotangle=78.73377](0.5420935,2.1538277)
	\psdots[dotsize=0.18,dotangle=78.73377](3.428318,-2.1938276)
      \pscircle[linewidth=0.03,dimen=outer,fillstyle=solid](2.0752058,0.06){0.4}
		\rput(2.06,0.045){I}
		\rput(4.8,2.325){${\cal O}(U/2)$}
		\rput(4.8,-2.435){${\cal O}(-U/2)$}
		\end{pspicture} 
		}
		\scalebox{0.8} 
		{
		\begin{pspicture}(0,-2.9)(2.455,0.55875)
\psline[linewidth=0.05cm,tbarsize=0.07055555cm 5.0,arrowsize=0.05291667cm 2.31,arrowlength=1.4,arrowinset=0.4]{|->}(0.1365625,-0.09125)(2.3365624,-0.09125)
		\rput(1.2040625,0.37375){tri-diagonalization}
		\rput(1.1779687,-0.38625){}
		\end{pspicture} 
		}
		\scalebox{0.8} 
		{
		\begin{pspicture}(0,-2.8)(8.782812,1.648125)
		\rput(5.65,0.7){$\varepsilon_0^+$}
		\rput(6.75,0.7){$\varepsilon_1^+$}
		\rput(7.55,0.7){$\varepsilon_l^+$}
		\rput(6.3,1.45){$t_0^+$}
	\psline[linewidth=0.02cm](3.6009376,-0.9503125)(5.6009374,1.0496875)
\pscircle[linewidth=0.03,dimen=outer,fillstyle=solid](4.6009374,0.0496875){0.4}
	\rput(4.5920315,0.0546875){I}
	\psdots[dotsize=0.2](3.6409376,-0.9103125)
	\psdots[dotsize=0.2](5.6409373,1.0896875)
	\psline[linewidth=0.02cm](0.8409375,-0.9103125)(1.8409375,-0.9103125)
	\psline[linewidth=0.02cm](2.6409376,-0.9103125)(3.6409376,-0.9103125)
	\psline[linewidth=0.02cm](5.6409373,1.0896875)(6.6409373,1.0896875)
	\psdots[dotsize=0.2](1.9209375,-0.9103125)
	\psdots[dotsize=0.2](7.5209374,1.0896875)
	\psdots[dotsize=0.2](2.7209375,-0.9103125)
	\psdots[dotsize=0.2](6.7209377,1.0896875)
\psline[linewidth=0.02cm,linestyle=dotted,dotsep=0.16cm](6.7409377,1.0896875)(7.4409375,1.0896875)
\psline[linewidth=0.02cm,linestyle=dotted,dotsep=0.16cm](1.9409375,-0.9103125)(2.6409376,-0.9103125)
	\psline[linewidth=0.02cm](7.6409373,1.0896875)(8.640938,1.0896875)
		\rput(3.6,-1.3){$\varepsilon_0^-$}
		\rput(2.7,-1.3){$\varepsilon_1^-$}
		\rput(1.9,-1.3){$\varepsilon_l^-$}
		\rput(3.3,-0.5){$t_0^-$}
		\rput(8.2,1.45){$t_l^+$}
		\rput(1.3,-0.5){$t_l^-$}
		\rput(4.8,0.9){$1/\sqrt{2}$}
		\rput(4.5,-0.7){$1/\sqrt{2}$}
		\end{pspicture} 
		}\caption{Mapping of the discretized SIAM onto 
two semi-infinite chains, coupled via the impurity. 
The states which have 
the energy $\xi_m=(U/2)$ and $\xi_m=-(U/2)$
in the atomic limit, respectively, 
are mapped onto the upper/lower chain.}\label{fig:siamtwochainmapping}
	\end{center}
\end{figure}
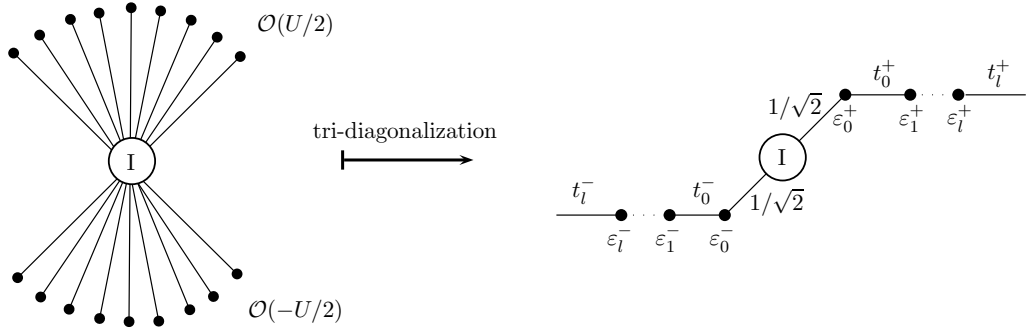

The two-chain mapping can be carried out technically along the lines
of the single-chain mapping (Lanczos tri-diagonalization~\cite{NRG}).
The transformed Hamiltonian reads
\begin{eqnarray}
\hat{H} &=& \hat{H}_0 + \hat{V} 
= \hat{H}_0 + \hat{V}_0 + \hat{V}_1 + \hat{V}_2
\label{eq:siamtwochain}\; , \\ 
\hat{H}_0&=& - \frac{U}{2} \sum_{l=0}^{(L-3)/2}\sum_{\sigma}
( \hat{\alpha}_{l, \sigma}^+ \hat{\alpha}_{l,\sigma} 
- \hat{\beta}_{l,\sigma}^+ \hat{\beta}_{l,\sigma} ) 
+ U (\hat{n}_{d,\uparrow}-1/2)(\hat{n}_{d,\downarrow}-1/2)\; , 
\label{eq:hnaught}\\
\hat{V}_0 &=& \sqrt{\frac{1}{2}} \sum_{\sigma} 
\big[  ( \hat{d}_{\sigma}^+ \hat{\alpha}_{0,\sigma} + 
\hat{d}_{\sigma}^+ \hat{\beta}_{0,\sigma} ) + \hbox{h.c.}\big]
\nonumber \; ,\\
\hat{V}_1 &=& \sum_{l=0}^{(L-3)/2}\sum_{\sigma} t_l 
\big[ ( \hat{\alpha}_{l,\sigma}^+\hat{\alpha}_{l+1,\sigma} + 
\hat{\beta}_{l,\sigma}^+ \hat{\beta}_{l+1,\sigma} ) +
\hbox{h.c.} \big] \; ,  \label{eq:perturbation} \\
\hat{V}_2 &=& \sum_{l=0}^{(L-3)/2}\sum_{\sigma} 
\varepsilon_{l} (\hat{\alpha}_{l,\sigma}^+\hat{\alpha}_{l,\sigma} 
- \hat{\beta}_{l,\sigma}^+ \hat{\beta}_{l,\sigma} ) \nonumber
\; .
\end{eqnarray}
The $\hat{\alpha}$-operators describe the electrons in the lower chain
 (lower Hubbard band)
and the $\hat{\beta}$-operators those in the upper chain (upper Hubbard band).
Due to particle-hole symmetry, the
electron-transfer amplitudes in the lower and upper chain are equal, 
$t_l^{-} =  t_l^{+}$, and the on-site energies in the 
lower and upper chains are opposite in sign,
$\varepsilon_l^{-}=-\varepsilon_l^{+}=\varepsilon_l-U/2$.
The mapping is shown in Fig.~\ref{fig:siamtwochainmapping}.
Later, we shall investigate the model in the strong-coupling limit.
Therefore, we separated the Hamiltonian into the starting Hamiltonian 
$\hat{H}_0$, eq.~(\ref{eq:hnaught}), 
and the perturbation $\hat{V}$, eq.~(\ref{eq:perturbation}). 
Note that $\hat{H}_0$ describes the atomic limit, $\hat{T}\equiv 0$, 
where there is no transfer between sites in the Hubbard model.

Our task is the calculation of the Green function on the impurity site,
eq.~(\ref{eq:whatweneedtocal}), for general on-site energies
$\varepsilon_l$ and electron-transfer parameters $t_l$.
In the two-chain geometry, these parameters assume the role of the
energies $\xi_m$ and the hybridizations $V_m$ in the star geometry.
Our approach relies on the order-by-order expansion
of all quantities in $1/U$,
\begin{eqnarray}
\varepsilon_l &=& 
\sum_{n=0}^{\infty} \varepsilon_l^{(n)} \left(\frac{1}{U}\right)^n \nonumber \;, \\
t_l &=&
\sum_{n=0}^{\infty} t_l^{(n)} \left(\frac{1}{U}\right)^n \;, 
\end{eqnarray}
for $l\geq 0$ whereby we implement the self-consistency 
equation~(\ref{what-we-need}).

\section{Kato--Takahashi perturbation theory}
\label{sec:KT-method}

In order to calculate the zero-temperature
Green function for the single-impurity Anderson model
in strong coupling, we adapt the Kato--Takahashi
perturbation theory.~\cite{kato,takahashi} 
The Kato--Takahashi perturbation
theory is particularly suitable for Hamiltonians $\hat{H}=\hat{H}_0+\hat{V}$
where the ground state of the unperturbed 
Hamiltonian~$\hat{H}_0$ is degenerate.

\subsection{General formalism}
\label{subsec:formalism}

\subsubsection{Transformation of the eigenstates}

The basic assumption of this perturbation theory is the existence of
an isometry $\Gamma_{i,N}$ from the $N$-particle 
eigenspace ${\cal E}_{i,N}^{(0)}$ of $H_0$ to the corresponding 
eigenspace ${\cal E}_{i,N}$ of $H$,
\begin{equation}
\Gamma_{i,N} : {\cal E}_{i,N}^{(0)} \rightarrow {\cal E}_{i,N} \; .
\end{equation}
Explicitly, the Kato--Takahashi projection 
operator~\cite{kato,takahashi,kalinowski,ruhl} 
is given by
\begin{equation}
\hat{\Gamma}_{i,N} = \hat{P}_{i,N} \hat{P}_{i,N}^{(0)} 
\bigl( \hat{P}_{i,N}^{(0)} \hat{P}_{i,N} \hat{P}_{i,N}^{(0)} \bigr)^{-1/2} \; .
\label{eq:defgamma}
\end{equation}
$\hat{\Gamma}_{i,N}$ is unitary provided we may identify 
the isomorphic subspaces ${\cal E}_{i,N}^{(0)}$ and ${\cal E}_{i,N}$.

The operators $\hat{P}_{i,N}^{(0)}$ project onto the $N$-particle eigenstates 
of $\hat{H}_0$,
i.e., states with energies $E_{i,N}^{(0)} = E_{0,N}^{(0)} + i U$. 
The operators $\hat{P}_{i,N}$ project onto the $N$-particle
eigenstates of $\hat{H}$.
They can be expressed as a perturbation series in terms of the
projectors $\hat{P}_{i,N}^{(0)}$ and the perturbation~$\hat{V}$,
\begin{eqnarray}
\hat{P}_{i,N} &=& \hat{P}_{i,N}^{(0)} 
+ \sum_{n=1}^{\infty} \hat{A}^{(n)}\; , \nonumber\\
\hat{A}^{(n)} &=& - \sum_{(n)} \tilde{S}^{k_1} \hat{V} 
\tilde{S}^{k_2} \cdots \hat{V} \tilde{S}^{k_{n+1}} \; ,
\label{eq:katoprojectionexpansion}
\end{eqnarray}
where we introduced the notations ($k\geq 1$)
\begin{eqnarray}
\tilde{S}^{0} = - \hat{P}_{i,N}^{(0)} &,&
\tilde{S}^k = \left(E_{i,N}^{(0)}-\hat{H}_0\right)^{-k}
\left(1 - P_{i,N}^{(0)}\right) 
\; ,\label{eq:katoabbrev} \\
\sum_{(l)} f(k_1, \ldots , k_m) &=& 
\mathop{\sum{}^{^{\scriptstyle \prime}}}_{k_1, \dots ,k_{m} =0}^l 
f(k_1, \ldots, k_m) 
\label{defprime} \; ,
\end{eqnarray}
and the prime on the sum in~(\ref{defprime})
implies $k_1+ \ldots + k_{m}=l$.
The square root operator in eq.~(\ref{eq:defgamma}) is defined 
from its series expansion,
\begin{equation}
\left(\hat{P}_{i,N}^{(0)} 
\hat{P}_{i,N} \hat{P}_{i,N}^{(0)}\right)^{-1/2}
=
\hat{P}_{i,N}^{(0)} +
\sum_{m=1}^{\infty} (-1)^m 
\frac{(2m-1)!!}{(2m)!!} 
\left[\sum_{n=1}^{\infty} \bar{A}^{(n)} \right]^m\; ,
\end{equation}
where $\bar{A}^{(n)}=\hat{P}_{i,N}^{(0)} \hat{A}^{(n)}\hat{P}_{i,N}^{(0)}$.
Up to and including fourth order in~$\hat{V}$ we find
\begin{equation}
\left(\hat{P}_{i,N}^{(0)} 
\hat{P}_{i,N} \hat{P}_{i,N}^{(0)}
\right)^{-1/2}
= 
\hat{P}_{i,N}^{(0)} -\frac{1}{2} \bar{A}^{(2)} -\frac{1}{2} \bar{A}^{(3)}
+ \left[ \frac{3}{8} \left(\bar{A}^{(2)} \right)^2 - \frac{1}{2} \bar{A}^{(4)} 
\right] + \ldots 
\; ,
\end{equation}
and corrections are of fifth order in the perturbation~$\hat{V}$.

Likewise, the Kato--Takahashi operator~(\ref{eq:defgamma})
can be obtained in a series expansion. 
Up to and including the third order in the perturbation~$\hat{V}$,
the Kato--Takahashi operator~$\hat{\Gamma}_{i,N}=
\sum_{n=0}^{\infty}\hat{\Gamma}_{i,N}^{(n)}$ reads
\begin{eqnarray}
\hat{\Gamma}_{i,N}^{(0)}&=& \hat{P}_{i,N}^{(0)} \nonumber \; ,\\
\hat{\Gamma}_{i,N}^{(1)}&=& \tilde{S}\hat{V}\hat{P}_{i,N}^{(0)}  \nonumber \; ,\\
\hat{\Gamma}_{i,N}^{(2)}&=& 
- \tilde{S}^{2}\hat{V}\hat{P}_{i,N}^{(0)}\hat{V}\hat{P}_{i,N}^{(0)}
-\frac{1}{2} \hat{P}_{i,N}^{(0)}\hat{V}\tilde{S}^{2}\hat{V}\hat{P}_{i,N}^{(0)}
+ \tilde{S}\hat{V}\tilde{S}\hat{V}\hat{P}_{i,N}^{(0)}  \; ,
\label{gammatothirdorder} \\
\hat{\Gamma}_{i,N}^{(3)}&=& 
\tilde{S}^{3}\hat{V}\hat{P}_{i,N}^{(0)}\hat{V}\hat{P}_{i,N}^{(0)}
\hat{V}\hat{P}_{i,N}^{(0)}
+\frac{1}{2} \hat{P}_{i,N}^{(0)}\hat{V}\tilde{S}^{3}\hat{V}\hat{P}_{i,N}^{(0)}
\hat{V}\hat{P}_{i,N}^{(0)}
+\frac{1}{2} \hat{P}_{i,N}^{(0)}\hat{V}\hat{P}_{i,N}^{(0)}\hat{V}
\tilde{S}^{3}\hat{V}\hat{P}_{i,N}^{(0)} \nonumber \\
&& -\tilde{S}^{2}\hat{V}\tilde{S}\hat{V}\hat{P}_{i,N}^{(0)}
\hat{V}\hat{P}_{i,N}^{(0)}
-\tilde{S}^{2}\hat{V}\hat{P}_{i,N}^{(0)}\hat{V}\tilde{S}\hat{V}\hat{P}_{i,N}^{(0)}
-\tilde{S}\hat{V}\tilde{S}^{2}\hat{V}\hat{P}_{i,N}^{(0)}\hat{V}\hat{P}_{i,N}^{(0)}
-\frac{1}{2} \hat{P}_{i,N}^{(0)}\hat{V}\tilde{S}^{2}\hat{V}
\tilde{S}\hat{V}\hat{P}_{i,N}^{(0)}
\nonumber \\
&& -\frac{1}{2} \tilde{S}\hat{V}\hat{P}_{i,N}^{(0)}\hat{V}\tilde{S}^{2}
\hat{V}\hat{P}_{i,N}^{(0)}
-\frac{1}{2} \hat{P}_{i,N}^{(0)}\hat{V}\tilde{S}\hat{V}
\tilde{S}^{2}\hat{V}\hat{P}_{i,N}^{(0)}
+ \tilde{S}\hat{V}\tilde{S}\hat{V}\tilde{S}\hat{V}\hat{P}_{i,N}^{(0)} 
\; .
\nonumber
\end{eqnarray}
With these definitions we can express the eigenstates 
of the Hamiltonian $\hat{H}$, $|\Psi\rangle \in {\cal E}_{0,L}$,
at half band-filling ($N=L$)
in terms of the eigenstates of the Hamiltonian $\hat{H}_0$,
$|\Phi\rangle \in {\cal E}_{0,L}^{(0)}$,
\begin{equation}
|\Psi\rangle = \hat{\Gamma}_{0,L} |\Phi\rangle \; .
\label{transformPhitoPsi}
\end{equation}
We introduce the state
\begin{equation}\label{eq:lanczosstartingvector}
|\Phi\rangle := \frac{1}{\sqrt{2}} \big( 
|\phi_{\uparrow}\rangle + |\phi_{\downarrow}\rangle  \bigr)\; ,
\end{equation}
which is a symmetric mixture of the two ground states
of the single-impurity model in the atomic limit at half band-filling,
\begin{eqnarray}
|\phi_{\uparrow}\rangle &=& 
\hat{d}_{\uparrow}^+ 
\prod_{l=0}^{(L-3)/2} \hat{\alpha}_{l,\uparrow}^+\hat{\alpha}_{l,\downarrow}^+
|\hbox{vac}\rangle 
\; ,
	\label{eq:groundstateuphnaught} \\
|\phi_{\downarrow}\rangle &=& 
\hat{d}_{\downarrow}^+ 
\prod_{l=0}^{(L-3)/2} \hat{\alpha}_{l,\uparrow}^+\hat{\alpha}_{l,\downarrow}^+
|\hbox{vac}\rangle 
	\label{eq:groundstatedownhnaught} \; .
\end{eqnarray}
Note that the overall phase of the states is fixed by the convention
that an electron on a particular site with spin~$\uparrow$ 
is placed to the left of an electron with spin~$\downarrow$.
Then, eq.~(\ref{eq:whatweneedtocal}) reduces to 
\begin{equation}\label{eq:gfsiamtransformed1}
G_{\rm LHB}(\omega<0) = \langle \Phi |
\hat{\Gamma}_{0,L}^+ \hat{d}_{\uparrow}^+ (\omega + \hat{H}- E_{0,L} 
- \imag \eta )^{-1} \hat{d}_{\uparrow} \hat{\Gamma}_{0,L}| \Phi\rangle \; , 
\end{equation}
which is an exact expression.
The average can now be taken in the known ground states of $\hat{H}_0$. 
The energy $E_{0,L}$ belongs to the symmetric
single-impurity Anderson model
and should not be confused with the ground-state energy $E_0(L)$
of the Hubbard model at half band-filling.

\subsubsection{Transformation of the multi-chain Hamiltonian}

In order to make use of eq.~(\ref{eq:gfsiamtransformed1}),
we must project $\hat{H}$ onto the eigenspaces of $\hat{H}_0$. 
Before we can continue, we must be aware of the fact 
that the lower Hubbard band consists of the primary lower Hubbard sub-band, 
centered at $(-U/2)$, and higher-order
sub-bands, centered at $(-U/2 -i U)$ ($i=1,2,\ldots)$. We assume that,
(i), these bands do not overlap in the Mott--Hubbard insulator, and that,
(ii), the degeneracies of the eigenspaces ${\cal E}_{i,N}^{(0)}$ 
are not lifted to all orders in perturbation theory, 
as is the case for the ground state at half band-filling.~\cite{kalinowski}
Under these assumptions, the operators $\hat{P}_{i,L-1}$ 
project onto those states, forming the $i$th lower sub-band. 
Since these projectors form a complete set,
$\sum_{i} \hat{P}_{i,L-1} = \openone$,
we may simplify the Green function~(\ref{eq:gfsiamtransformed1}) as
follows:
\begin{eqnarray}
G_{\rm LHB}(\omega) 
&=& 
\sum_{i,j} \langle \Phi|\hat{\Gamma}_{0,L}^{+} \hat{d}_{\uparrow}^+ 
\hat{P}_{i,L-1} \bigl(\omega + \hat{H}- E_{0,L} - \imag \eta \bigr)^{-1}
 \hat{P}_{j,L-1} \hat{d}_{\uparrow} \hat{\Gamma}_{0,L}|\Phi\rangle 
\nonumber \\
&=& \sum_{i} \langle\Phi|\hat{\Gamma}_{0,L}^{+} 
\hat{d}_{\uparrow}^+ \hat{P}_{i,L-1} 
\bigl(\omega + \hat{H}- E_{0,L} - \imag \eta \bigr)^{-1} 
\hat{P}_{i,L-1} \hat{d}_{\uparrow} \hat{\Gamma}_{0,L}|\Phi\rangle \; ,
\label{eq:split-one}
\end{eqnarray}
where we used the fact that 
the Hubbard sub-bands do not overlap. Now that
$\hat{\Gamma}_{i,N}^+\hat{\Gamma}_{i,N}= \hat{P}_{i,N}^{(0)}$ and
$\hat{\Gamma}_{i,N}\hat{\Gamma}_{i,N}^+= \hat{P}_{i,N}$ hold,
we may further simplify~(\ref{eq:split-one})
\begin{eqnarray}
G_{\rm LHB}(\omega) 
&=& \sum_{i} \langle \Phi | \hat{\Gamma}_{0,L}^+ \hat{d}_{\uparrow}^+ 
\hat{\Gamma}_{i,L-1}\hat{\Gamma}_{i,L-1}^+ 
\bigl(\omega + \hat{H} - E_{0,L} - \imag \eta \bigr)^{-1} 
\hat{\Gamma}_{i,L-1}\hat{\Gamma}_{i,L-1}^+ 
\hat{d}_{\uparrow} \hat{\Gamma}_{0,L}|\Phi\rangle 
\nonumber \\
&=&
\sum_{i} \langle \Phi| \tilde{d}_{i,\uparrow}^+ 
\bigl( \omega \hat{P}_{i,L-1}^{(0)} + \tilde{h}_{i,L-1} 
- E_{0,L}  \hat{P}_{i,L-1}^{(0)} 
- \imag \eta \bigr)^{-1} 
\tilde{d}_{i,\uparrow} |\Phi\rangle \label{eq:gfsiamtransformed2}
\end{eqnarray}
with the `reduced' operators
\begin{eqnarray}
\tilde{d}_{i,\sigma} = \hat{\Gamma}_{i,L-1}^{+} \hat{d}_{\sigma} 
\hat{\Gamma}_{0,L} &,& \tilde{d}^{+}_{i,\sigma} = 
\bigl( \tilde{d}_{i,\sigma} \bigr)^{+}, 
\label{eq:reducedcreator}
\\
\tilde{h}_{i,L-1} &=& \hat{\Gamma}_{i,L-1}^{+} \hat{H} \hat{\Gamma}_{i,L-1}
\label{eq:reducedhamiltonian} \; .
\end{eqnarray}
As seen from eq.~(\ref{eq:gfsiamtransformed2}), we need to work
with the `reduced Hamilton operator' $\tilde{h}_{i,L-1}$
which describes the dynamics of a hole in the symmetric
Anderson model.

Kato's perturbation theory~\cite{kato} also provides a perturbative
expression for the projected Hamiltonian,
\begin{eqnarray}
\left(\hat{H}-E_{i,N}^{(0)}\right) \hat{P}_{i,N} \hat{P}_{i,N}^{(0)} 
&=& \sum_{n=1}^{\infty} \tilde{B}^{(n)} \; , 
\label{katoexpandH}
\\
\tilde{B}^{(n)} &:=& - \sum_{(n-1)} \tilde{S}^{k_1}\hat{V}
\tilde{S}^{k_2}\hat{V} \ldots \tilde{S}^{k_n}V\hat{P}_{i,N}^{(0)} \; .
\end{eqnarray}
In order to evaluate~(\ref{eq:reducedhamiltonian}) we have to 
multiply the expression~(\ref{katoexpandH}) with 
$(\hat{P}_{i,N}^{(0)} \hat{P}_{i,N} \hat{P}_{i,N}^{(0)})^{-1/2}$
from the right and with $\hat{\Gamma}_{i,N}^{+}$ from the left.
The various orders in the interaction are then combined to give
\begin{equation}
\hat{\Gamma}_{i,N}^{+}\left(\hat{H}-E_{i,N}^{(0)}\right) 
\hat{\Gamma}_{i,N} = \sum_{n=0}^{\infty} \tilde{R}_{i,N}^{(n)}
\end{equation}
where, up to and including the third order in $1/U$, we find
\begin{eqnarray}
\tilde{R}_{i,N}^{(0)}
&=& \hat{P}_{i,N}^{(0)} \hat{V} \hat{P}_{i,N}^{(0)}\nonumber\\
\tilde{R}_{i,N}^{(1)}&=& \hat{P}_{i,N}^{(0)} 
\hat{V}\tilde{S}\hat{V}
\hat{P}_{i,N}^{(0)}
\nonumber \; ,\\
\tilde{R}_{i,N}^{(2)}&=& 
\hat{P}_{i,N}^{(0)} 
\hat{V}\tilde{S}\hat{V}\tilde{S}\hat{V}
\hat{P}_{i,N}^{(0)}
- \frac{1}{2} \left[
\hat{P}_{i,N}^{(0)} 
\hat{V}\tilde{S}^{2}\hat{V}\hat{P}_{i,N}^{(0)}\hat{V}
\hat{P}_{i,N}^{(0)}
+
\hat{P}_{i,N}^{(0)}\hat{V}\hat{P}_{i,N}^{(0)}\hat{V}\tilde{S}^{2}\hat{V}
\hat{P}_{i,N}^{(0)}
\right] 
\label{firstordersRtilde}
\; ,\\
\tilde{R}_{i,N}^{(3)}&=& 
\hat{P}_{i,N}^{(0)} 
\hat{V}\tilde{S}\hat{V}\tilde{S}\hat{V}\tilde{S}\hat{V}
\hat{P}_{i,N}^{(0)} 
+\frac{1}{2}\biggl[
\hat{P}_{i,N}^{(0)} 
\hat{V}\tilde{S}^{3}\hat{V}\hat{P}_{i,N}^{(0)}\hat{V}\hat{P}_{i,N}^{(0)}\hat{V}
\hat{P}_{i,N}^{(0)} 
+
\hat{P}_{i,N}^{(0)}
\hat{V}\hat{P}_{i,N}^{(0)}\hat{V}\hat{P}_{i,N}^{(0)} 
\hat{V}\tilde{S}^{3}\hat{V}
\hat{P}_{i,N}^{(0)}
\nonumber \\
&& -\hat{P}_{i,N}^{(0)} 
\hat{V}\tilde{S}^{2}\hat{V}\tilde{S}\hat{V}\hat{P}_{i,N}^{(0)}\hat{V}
\hat{P}_{i,N}^{(0)} 
-\hat{P}_{i,N}^{(0)} 
\hat{V}\tilde{S}^{2}\hat{V}\hat{P}_{i,N}^{(0)}\hat{V}\tilde{S}\hat{V}
\hat{P}_{i,N}^{(0)} 
-\hat{P}_{i,N}^{(0)} 
\hat{V}\tilde{S}\hat{V}\tilde{S}^{2}\hat{V}\hat{P}_{i,N}^{(0)}\hat{V}
\hat{P}_{i,N}^{(0)} 
\nonumber \\
&&-\hat{P}_{i,N}^{(0)}
\hat{V}\hat{P}_{i,N}^{(0)}\hat{V}\tilde{S}^{2}\hat{V}\tilde{S}\hat{V}
\hat{P}_{i,N}^{(0)}
-\hat{P}_{i,N}^{(0)}
\hat{V}\tilde{S}\hat{V}\hat{P}_{i,N}^{(0)}\hat{V}\tilde{S}^{2}\hat{V}
\hat{P}_{i,N}^{(0)}
-\hat{P}_{i,N}^{(0)}
\hat{V}\hat{P}_{i,N}^{(0)}\hat{V}\tilde{S}\hat{V}\tilde{S}^{2}\hat{V}
\hat{P}_{i,N}^{(0)}
\biggr]\; .
\nonumber
\end{eqnarray}
{}From the operators $\tilde{R}_{i,N}^{(n)}$ we readily obtain the
correction to the ground-state energy at half band-filling
($N=L$),
\begin{eqnarray}
E_{0,L}&=& E_{0,L}^{[-1]}
+ \sum_{n=0}^{\infty} E_{0,L}^{[n]} \nonumber \; , \\
E_{0,L}^{[n]} &=& \langle \phi_{\uparrow} |
\tilde{R}_{i,L}^{(n)} | \phi_{\uparrow} \rangle \; .
\label{ground-state-energy-correction}
\end{eqnarray}
Note that $E_{0,L}^{[-1]}=-(L-1)U/2-U/4$ because the $(L-1)/2$ sites in the
lower chain are doubly occupied in $|\phi_{\uparrow}\rangle$, the
impurity site is singly occupied, and the upper chain is empty,
cf.\ eqs.~(\ref{eq:hnaught}) and~(\ref{eq:groundstateuphnaught}).

In order to single out the leading-order contribution 
in eq.~(\ref{eq:gfsiamtransformed2}) we use
$E_{i,L-1}^{[-1]}-E_{0,L}^{[-1]}=U/2 + i U$
and define
\begin{equation}
\hat{{\cal L}}_{i,L-1} 
= \tilde{h}_{i,L-1} - \left(E_{0,L} + U/2+iU \right) 
\hat{P}_{i,L-1}^{(0)} 
= \sum_{n=0}^{\infty} \hat{{\cal L}}_{i,L-1}^{(n)}
\end{equation}
with
\begin{equation}
\hat{{\cal L}}_{i,L-1}^{(n)} =  \tilde{R}_{i,L-1}^{(n)} 
-E_{0,L}^{[n]}  \hat{P}_{i,L-1}^{(0)} \; .
\label{eq:lanczosoperator}
\end{equation}
Then, the Green function for the lower Hubbard 
band~(\ref{eq:gfsiamtransformed2})
can be written as the sum over the contributions
from the individual sub-bands around $\omega_i=-U/2-i U$,
\begin{equation}
\label{eq:gfsiamtransformed3}
G_{\rm LHB}(\omega) = 
\sum_{i} \langle \Phi | \tilde{d}_{i,\uparrow}^+ 
\left( (\omega + U/2 + i U) \hat{P}_{i,L-1}^{(0)} + \hat{\cal L}_{i,L-1} - 
\imag \eta \right)^{-1} \tilde{d}_{i,\uparrow} | \Phi\rangle \; .
\end{equation}
For the primary Hubbard sub-band, 
we will drop the index $i=0$ whenever possible.

\subsection{Matrix representation of the self-consistency equation}
\label{subsec:DMFT-selfconsistency}

The self-consistency equation~(\ref{what-we-need}) must be solved
for all frequencies in the respective sub-bands of the lower Hubbard band.
The density of states of the individual sub-bands can be viewed
as probability distributions which can be characterized by 
their moments. The idea is to express the Green function and the
hybridization function by their moments so that the 
self-consistency equation reduces to the condition that the two sets
of moments agree. In this way, only a countable set of numbers must
be compared. A suitable way to generate moments from a Green function
is provided by the Lanczos iteration procedure.

\subsubsection{Lanczos iteration}
For the starting vector $|0\rangle$ and the Hermitian operator $\hat{O}$, 
we use the following form of the Lanczos iteration:
\begin{eqnarray}
	|1\rangle &=& - \hat{O} |0\rangle + a_0 |0\rangle 
\label{eq:lanczos0}\;, \nonumber \\
	|n+1\rangle  &=& - \hat{O} |n\rangle  
+ a_n |n\rangle  + b_{n-1} |n-1\rangle \; , \quad n \geq 1\; , 
\label{eq:lanczosn}
\end{eqnarray}
where
\begin{eqnarray}
a_n &:=& \frac{\langle n | \hat{O} | n \rangle}{\langle n |n\rangle}
\; , \quad n \geq 0, \label{eq:lanczosenergy} \\
b_{n-1} &:=& \frac{\langle n-1| \hat{O} | n\rangle }{
\langle n-1| n-1\rangle } \equiv 
- \frac{\langle n| n\rangle }{\langle n-1| n-1\rangle } 
\; ,\quad n \geq 1 \label{eq:lanczoshopping} \; .
\end{eqnarray}
The matrix representation $\mathfrak{O}$ of $\hat{O}$ 
within the Lanczos basis $\left\{|n\rangle \right\}$ 
is tridiagonal, symmetric, and we have
\begin{eqnarray}
\mathfrak{O}_{n,n} &=& 
\frac{\langle n | \hat{O} | n \rangle}{\langle n |n\rangle}
= a_n \label{eq:matelements1}\; , \\
\mathfrak{O}_{n-1,n} &=& 
\frac{\langle n-1| \hat{O} | n\rangle }{
\sqrt{\langle n-1| n-1\rangle}\sqrt{\langle n| n\rangle} } =
- \sqrt{- b_n}  \; . \label{eq:matelements2}
\end{eqnarray}
In the following we use fracture letters for the matrix representations 
of the corresponding operators in their Lanczos basis. 
Note that the parameters $b_n$ can only be defined up to an arbitrary phase, 
which is a sign factor for real matrix elements.
Thus, the matrix $\mathfrak{O}'$, where we change the sign of 
an arbitrary off-diagonal element and of its symmetric counterpart,
represents the same operator $\hat{O}$. 

\subsubsection{Hybridization function}\label{ss:hybridizationfunction}

We introduce electron baths for every sub-band of the lower Hubbard band.
Starting from the single-impurity Anderson model in star geometry,
we write the hybridization function in the form
\begin{equation}
\Delta_{\rm LHB}(\omega)= 
\sum_{m} \sum_{i=0}^{\infty} 
\frac{ V_{i,m}^2}{(\omega+ U/2 + i U) -\xi_{i;m} - \imag \eta}
= \sum_{i=0}^{\infty} \Delta_i(\omega) \; .
\label{eq:delta-subbands}
\end{equation}
where $\Delta_i(\omega)$ denotes the contribution of the $i$th sub-band. 
We cast each $\Delta_i(\omega)$ into matrix form by applying the 
Lanczos iteration with the starting vector
\begin{equation}
|0\rangle_i:= \frac{1}{\sqrt{g_i}}\sum_{m} V_{i,m} \hat{a}_{i;m,\sigma}^+ 
|\hbox{vac}\rangle \; ,
\end{equation}
where $g_i$ is the weight of the $i$th sub-band in the density of states,
$\sum_i g_i=1/2$, see appendix~\ref{appD}.
With the (discretized) Hamiltonian for the bath electrons in star geometry
\begin{equation}
\hat{H}_{\Delta,i}= \sum_{m,\sigma} 
\xi_{i,m} \hat{a}_{i;m,\sigma}^+\hat{a}_{i;m,\sigma}
\end{equation}
we may write
\begin{eqnarray}
\Delta_i(\omega) &=& 
{}_i \langle 0 |
 \left( \left( \omega + U/2 + i U\right) 
- \hat{H}_{\Delta,i} -\imag \eta \right)^{-1} \, |0\rangle_i
\label{eq:hybrid1} \\ 
&\equiv& 
\left(\left( \left(\omega + U/2 + i U\right) \openone - 
\mathfrak{h}_{\Delta,i} - \imag \eta\right)^{-1} \right)_{00} \; . 
\label{eq:hybrid2}
\end{eqnarray}
This form can be verified by noting 
that $[\hat{H}_{\Delta,i}]^n |0\rangle_i 
= 1/\sqrt{g_i} \sum_{m,\sigma} V_{i,m} (\xi_{i,m})^{n} \hat{a}_{i;m,\sigma}^+
|\hbox{vac}\rangle$.

We note that the mapping of the single-impurity model
from the star geometry to the multi-chain geometry is based
on the Lanczos procedure.~\cite{NRG} Therefore,
the starting vector $|0\rangle_i$ is identical to an electron
at the first site, 
$|0\rangle_i=(1/\sqrt{2})\hat{\alpha}_{i;0,\sigma}^+|\hbox{vac}\rangle$, of
the $i$th lower chain in the multi-chain geometry. 
Thus, the Hamiltonian $\hat{H}_{\Delta,i}$ for the bath electrons 
can also be written in the form
\begin{equation}
\hat{H}_{\Delta,i} := \sum_{l=0}^{\infty}\sum_{\sigma}
\left\{
t_{i;l}\left(\hat{\alpha}_{i;l,\sigma}^+ \hat{\alpha}_{i;l+1,\sigma} + \hbox{h.c.}
\right) + \varepsilon_{i;l} \hat{\alpha}_{i;l,\sigma}^+ \hat{\alpha}_{i;l,\sigma} 
\right\} \; .
\end{equation}
Then, the matrix $\mathfrak{h}_{\Delta,i}$ representing $\hat{H}_{\Delta,i}$ 
in the Lanczos basis reads
\begin{equation}\label{eq:matrixforhdelta}
\mathfrak{h}_{\Delta,i}=
\left(
\begin{array}{cccccc}
		\varepsilon_{i;0} & t_{i;0} &  &  & &\\
		t_{i;0} & \varepsilon_{i;1} & t_{i;1} & & & \\
		& t_{i;1} & \varepsilon_{i;2} & t_{i;2} & & \\
	& &  \ddots & \ddots & \ddots & \\
	\end{array} 
\right)\; , 
\end{equation}
where the entries not shown are zero.
The parameters $\varepsilon_{i;m}$ and $t_{i;m}$ define the single-impurity 
Anderson model in its multi-chain geometry.

\subsubsection{Green function}
\label{ss:greenfunction}
For the Green function~(\ref{eq:gfsiamtransformed3})
we use 
\begin{equation}
\label{startlanczosgreen}
|\Psi_0\rangle=\tilde{d}_{\uparrow} |\Phi\rangle
\end{equation}
as the starting vector, see eq.~(\ref{eq:lanczosstartingvector}),
and 
\begin{equation}
\hat{O} = \hat{\cal L}_{i,L-1}
\label{operatorlanczosgreeen}
\end{equation}
as the operator in the Lanczos iteration, see eq.~(\ref{eq:lanczosoperator}).
In this way we obtain the matrix representation
\begin{equation}\label{eq:greenfunctionmatrixrepresentation}
G_{\rm LHB}(\omega)= 
\sum_{i=0}^{\infty}
\left(\left( \left(\omega + U/2 + i U\right) \openone 
+\mathfrak{L}_i  - \imag \eta\right)^{-1} \right)_{00} \; , 
\end{equation}
where the structure of $\mathfrak{L}_i$ is given by
\begin{equation}
\label{eq:matelem3}
	\mathfrak{L}_{i}=\left(
\begin{array}{cccccc}
		e_{i;0} & \tau_{i;0} &  &  & &\\
		\tau_{i;0} & e_{i;1} & \tau_{i;1} & & &  \\
		& \tau_{i;1} & e_{i;2} & \tau_{i;2} & &  \\
    &	 &  \ddots & \ddots & \ddots & \\
	\end{array}
\right)\; .
\end{equation}
The parameters $e_{i;m}$ and $\tau_{i;m}$
must be calculated 
from eqs.~(\ref{eq:matelements1}) and~(\ref{eq:matelements2}).

\subsubsection{Self-consistency equation}
\label{ss:selfconsistencymatrix}
For $t\equiv 1$ as our unit of energy, 
the self-consistency equation~(\ref{what-we-need}) reads
($\omega<0$)
\begin{equation}
\sum_{i=0}^{\infty}
\left(\left( \left(\omega + U/2 + i U\right) \openone - 
\mathfrak{h}_{\Delta,i} - \imag \eta\right)^{-1} \right)_{00}
=
\sum_{i=0}^{\infty}
\left(\left( \left(\omega + U/2 + i U\right) \openone 
+\mathfrak{L}_i  - \imag \eta\right)^{-1} \right)_{00} \; .
\end{equation}
In this work we are mainly interested in the primary lower Hubbard band.
As we show in appendix~\ref{appD}, it is the only lower Hubbard band
with non-vanishing spectral weight up to and including 
third order in $1/U$.
Up to this order, we may therefore write 
\begin{equation}\label{eq:selfconsistencylhb1}
\left(\left( \left(\omega + U/2 \right) \openone - 
\mathfrak{h}_{\Delta} - \imag \eta\right)^{-1} \right)_{00}
=
\left(\left( \left(\omega + U/2 \right) \openone 
+\mathfrak{L}  - \imag \eta\right)^{-1} \right)_{00} \; ,
\end{equation}
where we have dropped the subscript $i=0$.
Reckoning with~(\ref{eq:selfconsistencylhb1}), we realize that
\begin{equation}\label{eq:selfconsistencylhbmatrix1}
\mathfrak{h}_{\Delta} = - \mathfrak{L}
\end{equation}
is a sufficient condition to ensure the self-consistency. From
the continued fraction expansion of the hybridization function
and the Green function it can readily be shown that
it also is a necessary condition. Therefore, the self-consistency condition
reduces to
\begin{equation}
\label{eq:onlynumbersremain}	
\varepsilon_n = - e_n \quad \hbox{and} \quad |t_n| = |\tau_n| \; .
\end{equation}
We already remarked that 
the off-diagonal Lanczos parameters are only defined up to a sign factor.
Note that eq.~(\ref{eq:onlynumbersremain})
is a vast simplification over~(\ref{what-we-need})
because, due to the matrix structure, 
we only have to equate numbers and not functions. 
However, since all our calculations are implicitly 
done in the thermodynamic limit, 
there is a (countably) infinite set of parameters to fix.
As we shall show explicitly up to third order in $1/U$, 
the locality of the Hubbard interaction guarantees
that there is an index $l_m$ in $m$th 
order perturbation theory such that the Lanczos parameters $\tau_{n}$ 
and $e_{n}$ become constant for $n\geq l_m$.

\section{Solution of the DMFT equation}
\label{sec:results}

In the first part of this section we show how the DMFT equation 
is solved to leading order in $1/U$. 
In the second part we summarize the results to third order.

\subsection{Calculations to leading order}
\label{subsec:leading-order}

First, we calculate the ground-state energy and 
set up the starting vector for the Lanczos iteration.
Next, we calculate the action of the Lanczos operator
on the states with a single hole in the lower Hubbard chain. 
Then, we derive the parameters $\varepsilon_{0}^{(0)}$,
$\varepsilon_{1}^{(0)}$, $t_{0}^{(0)}$, and $t_{1}^{(0)}$
from the first two Lanczos iterations.
Lastly, we prove the result $\varepsilon_l^{(0)}=0$ and $t_l^{(0)}=1$ 
for all~$l$ by induction.

\subsubsection{Ground-state energy and starting vector for the Lanczos iteration}

To leading order, we have $\hat{\Gamma}_{L}^{(0)}=\hat{P}_{L}^{(0)}$ 
from~(\ref{gammatothirdorder}). From~(\ref{transformPhitoPsi}) 
we see that the ground state at half band-filling
is not transformed to leading order, $|\Psi^{(0)}\rangle=
|\Phi\rangle$, see eqs.\ 
(\ref{eq:lanczosstartingvector})-(\ref{eq:groundstatedownhnaught}). 
The correction to the ground-state energy to leading order 
follows from eq.~(\ref{ground-state-energy-correction}) as
\begin{equation}
E_{0,L}^{[0]} = \langle \phi_{\uparrow} |
\tilde{R}_{L}^{(0)} | \phi_{\uparrow} \rangle 
= \langle \phi_{\uparrow} | \hat{V} | \phi_{\uparrow} \rangle 
= \langle \phi_{\uparrow} | \hat{V}_2 | \phi_{\uparrow} \rangle 
= 2\sum_{l=0}^{(L-3)/2}\varepsilon_l \; .
\end{equation}
The operators $\hat{V}_0$ and $\hat{V}_1$ do not contribute
because they  modify $|\phi_{\uparrow}\rangle$.
The contribution of $\hat{V}_2$ is readily calculated because
the sites of the lower chain are doubly occupied, see 
eq.~(\ref{eq:perturbation}).

According to~(\ref{startlanczosgreen}),
the starting vector for the Lanczos iteration
is given by
\begin{eqnarray}
|\Psi^{(0)}_0\rangle &=& \hat{P}_{L-1}^{(0)}\hat{d}_{\uparrow} 
\hat{P}_{L}^{(0)}|\Phi\rangle \equiv |\phi_{-1}\rangle \nonumber \; ,\\
|\phi_{-1}\rangle &=& \sqrt{\frac{1}{2}}
\hat{d}_{\uparrow}| \phi_{\uparrow}\rangle = \sqrt{\frac{1}{2}}
\prod_{l=0}^{(L-3)/2} \hat{\alpha}_{l,\uparrow}^+\hat{\alpha}_{l,\downarrow}^+
|\hbox{vac}\rangle 
\; .
\label{defphiminusone}
\end{eqnarray}
Note that, in general, the starting vector is not normalized to unity.

\subsubsection{Lanczos operator}
The operator for the Lanczos iteration~(\ref{operatorlanczosgreeen})
is given by
 \begin{equation}
\hat{\cal L}_{L-1}^{(0)}
=\tilde{R}_{L-1}^{(0)} - E_{0,L}^{[0]}\hat{P}_{L-1}^{(0)} 
= \hat{P}_{L-1}^{(0)}\left(\hat{V}- E_{0,L}^{[0]}\right)\hat{P}_{L-1}^{(0)} 
\; ,
\end{equation}
see eqs.~(\ref{firstordersRtilde}) and~(\ref{eq:lanczosoperator}).
In the course of the calculations, we shall need the 
eigenbasis of $\hat{P}_{L-1}^{(0)}$, i.e., single-hole states
in the half-filled ground states of $\hat{H}_0$.
Apart from $|\phi_{-1}\rangle$ in~(\ref{defphiminusone}),
we define for $n\geq 0$ (see eqs.~(\ref{eq:groundstateuphnaught}) 
and~(\ref{eq:groundstatedownhnaught}))
\begin{eqnarray}
|\phi_{n;u}\rangle &:=& 
\sqrt{\frac{1}{2}}  \hat{d}_{\downarrow}^+\hat{\alpha}_{n,\downarrow}
\prod_{l=0}^{(L-3)/2} 
\hat{\alpha}_{l,\uparrow}^+\hat{\alpha}_{l,\downarrow}^+
|\hbox{vac}\rangle
\; ,
\nonumber\\
|\phi_{n;d}\rangle &:=& -\sqrt{\frac{1}{2}} 
\hat{d}_{\uparrow}^+\hat{\alpha}_{n,\uparrow}
\prod_{l=0}^{(L-3)/2} 
\hat{\alpha}_{l,\uparrow}^+\hat{\alpha}_{l,\downarrow}^+
|\hbox{vac}\rangle
\; ,
\label{single-hole-firstset}
\\
|\chi_{n}\rangle &:=& 
 -\sqrt{\frac{1}{2}} 
\hat{d}_{\downarrow}^+\hat{\alpha}_{n,\uparrow}
\prod_{l=0}^{(L-3)/2} 
\hat{\alpha}_{l,\uparrow}^+\hat{\alpha}_{l,\downarrow}^+
|\hbox{vac}\rangle
\; ,\nonumber
\end{eqnarray}
and their useful linear combinations
\begin{eqnarray}
|\gamma_{n}\rangle &:=& (-1)^n \sqrt{\frac{1}{2}} 
\left( |\phi_{n;u}\rangle-|\phi_{n;d}\rangle\right)
\; ,\nonumber\\
|m_{n;u}\rangle &:=& (-1)^n \sqrt{\frac{1}{2}} 
\left( |\phi_{n;u}\rangle+|\chi_{n}\rangle\right)
\label{single-hole-secondset}\; ,\\
|m_{n;d}\rangle &:=& (-1)^n \sqrt{\frac{1}{2}} 
\left( |\phi_{n;d}\rangle+|\chi_{n}\rangle\right)
\; . \nonumber
\end{eqnarray}
Note that the states are not normalized but they are site-orthogonal
in the sense that the overlap between states with different site indices
is zero.

The action of the Lanczos operator $\hat{\cal L}_{L-1}^{(0)}$
on the states is readily calculated. One finds
for $n=0$,
\begin{eqnarray}
\hat{\cal L}_{L-1}^{(0)} |\phi_{-1}\rangle &=& |\gamma_0\rangle
\; , \nonumber \\
\hat{\cal L}_{L-1}^{(0)} |\gamma_0\rangle &=& 
|\phi_{-1}\rangle-\varepsilon_0^{(0)} |\gamma_0\rangle+t_0^{(0)}|\gamma_1\rangle
\; , \nonumber \\
\hat{\cal L}_{L-1}^{(0)} |m_{0;u}\rangle &=&
\frac{1}{2} |\phi_{-1}\rangle -\varepsilon_0^{(0)} |m_{0;u}\rangle 
+t_0^{(0)}|m_{1;u}\rangle
\; , \nonumber \\
\hat{\cal L}_{L-1}^{(0)} |m_{0;d}\rangle &=&
-\frac{1}{2} |\phi_{-1}\rangle -\varepsilon_0^{(0)} |m_{0;d}\rangle 
+t_0^{(0)}|m_{1;d}\rangle
\; , \label{nzeroaction}
\end{eqnarray}
and, for $n\geq 1$ and $x_n=\gamma_n,m_{n;u},m_{n;d}$,
\begin{equation}
\hat{\cal L}_{L-1}^{(0)} |x_n\rangle =
t_{n-1}^{(0)} |x_{n-1}\rangle -\varepsilon_n^{(0)}|x_n\rangle 
+t_n^{(0)}|x_{n+1}\rangle \; .
\label{n-one-action}
\end{equation}
The effect of the Lanczos operator is identical for all $n\geq 1$.
Note that this holds true in $m$th-order perturbation theory
in $1/U$ for $n\geq m+1$.

\subsubsection{First and second Lanczos iterations}

In the first Lanczos iteration we must determine the state
\begin{equation}
|\Psi_1^{(0)}\rangle := -\hat{\cal L}_{L-1}^{(0)} |\Psi_0^{(0)}\rangle
+e_0^{(0)} |\Psi_0^{(0)}\rangle
\end{equation}
with
\begin{equation}
e_0^{(0)} = \frac{ \langle \Psi_0^{(0)}| \hat{\cal L}_{L-1}^{(0)} 
|\Psi_0^{(0)}\rangle}{\langle \Psi_0^{(0)}| \Psi_0^{(0)}\rangle}
\; .
\end{equation}
With the help of (\ref{nzeroaction}) we find
\begin{equation}
|\Psi_1^{(0)}\rangle = -| \gamma_0\rangle \quad , \quad e_0^{(0)}=0 
\end{equation}
because $e_0^{(0)}=2 \langle \phi_{-1} | \gamma_0\rangle=0$.
The self-consistency equation~(\ref{eq:onlynumbersremain})
then gives $\varepsilon_0^{(0)}=0$. Furthermore, we have
\begin{equation}
\tau_0^{(0)} = 
\frac{ \langle \Psi_0^{(0)}| \hat{\cal L}_{L-1}^{(0)} 
|\Psi_1^{(0)}\rangle}{\langle \Psi_0^{(0)}| \Psi_0^{(0)}\rangle}
=-2 \langle \gamma_0|\gamma_0\rangle =-1\; , 
\end{equation}
so that $t_0^{(0)}=1$ follows from 
the self-consistency equation~(\ref{eq:onlynumbersremain}).

In the second iteration we can use the results from the first iteration.
We calculate
\begin{equation}
|\Psi_2^{(0)}\rangle := -\hat{\cal L}_{L-1}^{(0)} |\Psi_1^{(0)}\rangle
+e_1^{(0)} |\Psi_1^{(0)}\rangle +\tau_0^{(0)}  |\Psi_0^{(0)}\rangle
\end{equation}
with
\begin{equation}
e_1^{(0)} = \frac{ \langle \Psi_1^{(0)}| \hat{\cal L}_{L-1}^{(0)} 
|\Psi_1^{(0)}\rangle}{\langle \Psi_1^{(0)}| \Psi_1^{(0)}\rangle }
=2\left(\langle \gamma_0|\phi_{-1}\rangle 
+ \langle \gamma_0|\gamma_{1}\rangle\right) =0 \; ,
\end{equation}
where we used $\varepsilon_0^{(0)}=0$ and $t_0^{(0)}=1$ in~(\ref{nzeroaction}).
The self-consistency equation~(\ref{eq:onlynumbersremain})
then gives $\varepsilon_1^{(0)}=0$. 

The second state in the Lanczos iteration reduces to
\begin{equation}
|\Psi_2^{(0)}\rangle = |\gamma_1\rangle \;,
\end{equation}
and we find
\begin{equation}
\tau_1^{(0)} = 
\frac{ \langle \Psi_1^{(0)}| \hat{\cal L}_{L-1}^{(0)} 
|\Psi_2^{(0)}\rangle}{\langle \Psi_1^{(0)}| \Psi_1^{(0)}\rangle}
=-2 \left(\langle \phi_{-1} |\gamma_1\rangle +
\langle \gamma_1|\gamma_1\rangle \right) =-1\; , 
\end{equation}
so that $t_1^{(0)}=1$ follows from 
the self-consistency equation~(\ref{eq:onlynumbersremain}).

\subsubsection{Induction}

Now we are in the position to prove by induction that
$\varepsilon_n^{(0)}=e_n^{(0)}=0$ and $\tau_n^{(0)}=-1=-t_n^{(0)}$.
Let this assumption be true for $1\leq n\leq M-2$ ($M\geq 3$) and 
assume for $1\leq n\leq M-1$ ($M\geq 3$) that
\begin{equation}
|\Psi_n^{(0)}\rangle = (-1)^n |\gamma_{n-1}\rangle \; .
\end{equation}
Then, we calculate
\begin{equation}
|\Psi_M^{(0)}\rangle := -\hat{\cal L}_{L-1}^{(0)} |\Psi_{M-1}^{(0)}\rangle
+e_{M-1}^{(0)} |\Psi_{M-1}^{(0)}\rangle +\tau_{M-2}^{(0)}  |\Psi_{M-2}^{(0)}\rangle
\end{equation}
with
\begin{equation}
e_{M-1}^{(0)} = \frac{ \langle \Psi_{M-1}^{(0)}| \hat{\cal L}_{L-1}^{(0)} 
|\Psi_{M-1}^{(0)}\rangle}{\langle \Psi_{M-1}^{(0)}| \Psi_{M-1}^{(0)}\rangle }
=2(-1)^{M-1+M}\left(\langle \gamma_{M-2}|\gamma_{M-3}\rangle 
+ \langle \gamma_{M-2}|\gamma_{M-1}\rangle\right) =0 \; ,
\end{equation}
where we used $\varepsilon_{M-2}^{(0)}=0$ and $t_{M-2}^{(0)}=t_{M-3}^{(0)}=1$ 
in~(\ref{n-one-action}).
The self-consistency equation~(\ref{eq:onlynumbersremain})
then gives $\varepsilon_{M-1}^{(0)}=0$ which proves the next step
in the induction for $\varepsilon_n^{(0)}$.

The next state in the Lanczos iteration reduces to
\begin{equation}
|\Psi_{M}^{(0)}\rangle = (-1)^M\left(
|\gamma_{M-3}\rangle + |\gamma_{M-1}\rangle
\right) - (-1)^{M-2}|\gamma_{M-3}\rangle =
(-1)^M |\gamma_{M-1}\rangle \;,
\end{equation}
which proves the next step
in the induction for $|\Psi_n^{(0)}\rangle$.

Finally, we find
\begin{equation}
\tau_{M-1}^{(0)} = 
\frac{ \langle \Psi_{M-1}^{(0)}| \hat{\cal L}_{L-1}^{(0)} 
|\Psi_M^{(0)}\rangle}{\langle \Psi_{M-1}^{(0)}| \Psi_{M-1}^{(0)}\rangle}
=-2(-1)^{M+M} \left(\langle \gamma_{M-3} |\gamma_{M-1}\rangle +
\langle \gamma_{M-1}|\gamma_{M-1}\rangle \right) =-1\; , 
\end{equation}
so that $t_{M-1}^{(0)}=1$ follows from 
the self-consistency equation~(\ref{eq:onlynumbersremain}).
This proves the next step in the induction for $t_n^{(0)}$.

We recall that our approach strongly relies on the simple 
form~(\ref{what-we-need}) of the self-consistency equation.
For a general form of the bare density of states $\rho(\omega)$,
the calculation of the density of states for the lower Hubbard 
band to leading order is a demanding task.~\cite{MVS}

\subsection{Results up to third order}

The calculations up to third order are straightforward but tedious.
Simplifications arise from the fact that we are interested
in the half-filled case ($N=L$) and the situation with a single hole
($N=L-1$). Moreover, the results to leading order simplify
the analysis considerably. Details are given in Ref.~[\onlinecite{ruhl}].
Here we summarize the results.

\subsubsection{Ground-state energy and starting vector for the Lanczos iteration}

The ground-state energy is given by
\begin{equation}
E_{0,L}+U(L-1) +\frac{U}{4} =
\langle \phi_{\uparrow}| \tilde{R}_{L}|\phi_{\uparrow}\rangle =
-\frac{1}{U}-\frac{3}{2U^3} +\sum_{l=0}^{(L-3)/2} \varepsilon_l 
+ {\cal O}\left(\frac{1}{U^4}\right) \; .
\label{energythirdorder}
\end{equation}
The starting vector for the Lanczos iteration
to third order is given by
\begin{equation}
|\Psi_0\rangle = |\phi_{-1} \rangle
-\frac{1}{U}|m_{0;u}\rangle +
\frac{1}{U^2} \left( -\frac{1}{2} |\phi_{-1} \rangle
+ |m_{1;u}\rangle \right)
-\frac{1}{U^3} \left( \frac{7}{4}  |m_{0;u}\rangle 
+ |m_{2;u}\rangle \right) \; .
\label{startingthirdorder}
\end{equation}
In the actual derivation, the results of the lowest-order 
calculations are used in first order, those of the first-order
calculations are required in second order, and so on.

\subsubsection{Lanczos operator}

Up to and including the third order in $1/U$ the Lanczos operator 
$\hat{\cal L}_{L-1}\equiv \hat{\cal L}$ reads
\begin{eqnarray}
\hat{\cal L}
&=& \hat{P}_{L-1}^{(0)}
\biggl(\hat{V}+\frac{1}{U}+\frac{3}{2U^3} -\sum_{l=0}^{(L-3)/2} \varepsilon_l
\biggr) \hat{P}_{L-1}^{(0)} \nonumber \\
&&-\frac{1}{U} \hat{h_1} +\frac{1}{U^2}\left(
\hat{h}_2-\frac{1}{2}\hat{h}_1\hat{h}_0- \frac{1}{2}\hat{h}_0\hat{h}_1
\right)
\\
&&+\frac{1}{U^3}\left(
-\hat{h}_3+\hat{h}_2\hat{h}_0+\hat{h}_0\hat{h}_2
-\frac{1}{2}\hat{h}_1(\hat{h}_0)^2- \frac{1}{2}(\hat{h}_0)^2\hat{h}_1
+(\hat{h}_1)^2
\right) \; .
\nonumber 
\end{eqnarray}
Here, we used the abbreviations
\begin{eqnarray}
\hat{h}_0&=& \hat{P}_{L-1}^{(0)} \bar{V}_0\hat{P}_{L-1}^{(0)}
\nonumber \; ,\\
\hat{h}_1&=& \hat{P}_{L-1}^{(0)} \hat{V}_0\hat{S}\hat{V}_0\hat{P}_{L-1}^{(0)}
\nonumber \; ,\\
\hat{h}_2&=& \hat{P}_{L-1}^{(0)} \hat{V}_0\hat{S}
\bar{V}_0
\hat{S}\hat{V}_0\hat{P}_{L-1}^{(0)}
\nonumber \; ,\\
\hat{h}_3&=& \hat{P}_{L-1}^{(0)} \hat{V}_0\hat{S}
\bar{V}_0\hat{S}\bar{V}_0
\hat{S}\hat{V}_0\hat{P}_{L-1}^{(0)}
\; ,
\end{eqnarray}
where $\bar{V}_0=\hat{V}_0+\hat{V}_1^{(0)}$ describes the
coupling of the chains to the impurity with amplitude $V_0=1/\sqrt{2}$
and the free hole motion along the chain with amplitude $t_l=-1$.
The operator $\hat{S}$ measures the inverse number of excitations
above the ground states of $\hat{H}_0$ for $N=L-1$ particles, 
\begin{equation}
\hat{S}=\sum_{j=1}^{\infty}\frac{\hat{P}_{j,L-1}^{(0)}}{j} \;.
\end{equation}
The remaining task is the calculation of the action of the Lanczos
operator on the single-hole states~(\ref{single-hole-firstset}) and
(\ref{single-hole-secondset}).

Up to and including third order in $1/U$ we find 
for the state with the hole at the impurity
\begin{equation}
\hat{\cal L} |\phi_{-1}\rangle 
= \left(1-\frac{3}{4U^2} \right)
|\gamma_0\rangle
+\left( \frac{1}{U} + \frac{3}{2U^3} \right)|\phi_{-1}\rangle 
+ \frac{1}{U^3}|\gamma_1\rangle \; .
\label{action-one}
\end{equation}
For the states with the hole at site~$n$ we find for $n=0$
\begin{eqnarray}
\hat{\cal L} |\gamma_0\rangle &=&
\left(1-\frac{3}{4U^2}\right)|\phi_{-1}\rangle 
- \left(\frac{1}{2U} +\frac{3}{2U^3}
+\sum_{l=0}^3\frac{\varepsilon_0^{(l)}}{U^l}\right)| \gamma_0\rangle 
+ \left(\frac{1}{4U^2}+\sum_{l=0}^3\frac{t_0^{(l)}}{U^l}\right)|\gamma_1\rangle
\nonumber \\
&&  -\frac{1}{4U^3} |\gamma_2\rangle \; ,
\label{action-two}\\
\hat{\cal L}|m_{0;u}\rangle &=&
\frac{1}{2}\left(1-\frac{3}{4U^2}\right) |\phi_{-1}\rangle 
+\frac{1}{2U} |m_{0;d}\rangle - \frac{1}{4U^2} |m_{1;d}\rangle 
- \sum_{l=0}^2\frac{\varepsilon_0^{(l)}}{U^l}| m_{0;u}\rangle 
+ \sum_{l=0}^2\frac{t_0^{(l)}}{U^l}|m_{1;u}\rangle 
\; , \nonumber \\
\hat{\cal L}|m_{0;d}\rangle &=&
-\frac{1}{2}\left(1-\frac{3}{4U^2}\right) |\phi_{-1} \rangle
+\frac{1}{2U} |m_{0;u}\rangle - \frac{1}{4U^2} |m_{1;u}\rangle  
- \sum_{l=0}^2\frac{\varepsilon_0^{(l)}}{U^l}| m_{0;d}\rangle 
+ \sum_{l=0}^2\frac{t_0^{(l)}}{U^l}|m_{1;d}\rangle 
\; .\nonumber
\end{eqnarray}
For $n\geq 1$ we find
\begin{eqnarray}
\hat{\cal L}|\gamma_n\rangle &=&
\delta_{n,1}\frac{1}{4U^2}|\gamma_{0}\rangle 
+\delta_{n,1}\frac{1}{U^3}|\phi_{-1}\rangle
-\delta_{n,2}\frac{1}{4U^3}|\gamma_0\rangle 
\nonumber\\
&& 
+\sum_{l=0}^3\frac{1}{U^l} \left(
t_{n-1}^{(l)}|\gamma_{n-1}\rangle 
- \varepsilon_n^{(l)}|\gamma_n\rangle 
+t_{n}^{(l)}|\gamma_{n+1}\rangle \right)
\; , \nonumber \\
\hat{\cal L}|m_{n;u}\rangle &=&
-\delta_{n,1}\frac{1}{4U^2}|m_{0;d}\rangle
+\sum_{l=0}^2\frac{1}{U^l} \left(
t_{n-1}^{(l)}|m_{n-1;u}\rangle 
- \varepsilon_n^{(l)}|m_{n;u}\rangle 
+t_{n}^{(l)}|m_{n+1;u}\rangle \right)
\; ,\label{action-three}\\
\hat{\cal L}|m_{n;d}\rangle &=&
-\delta_{n,1}\frac{1}{4U^2}|m_{0;u}\rangle
+\sum_{l=0}^2\frac{1}{U^l} \left(
t_{n-1}^{(l)}|m_{n-1;d}\rangle - \varepsilon_n^{(l)}|m_{n;d}\rangle 
+ t_{n}^{(l)}|m_{n+1;d}\rangle \right)
\; . \nonumber 
\end{eqnarray}
In the above formulae, we did not include the third-order contributions
to $|m_{n;u}\rangle$ and $|m_{n;d}\rangle$ because
they are not needed for the third-order calculations.

With the help of the starting vector~(\ref{startingthirdorder})
and the action of the Lanczos operator~(\ref{action-one})-(\ref{action-three}),
the Lanczos vectors can be generated iteratively, along
with the values for $e_l$ and $\tau_l$, see eqs.~(\ref{eq:matelements1}),
(\ref{eq:matelements2}), and (\ref{eq:matelem3}).

\subsubsection{First and second Lanczos iterations}

The first and second Lanczos iterations give the vectors
\begin{eqnarray}
|\Psi_1\rangle &=& -|\gamma_0\rangle 
+\frac{1}{U} \left(-\frac{1}{2}|\phi_{-1}\rangle +|m_{1;u}\rangle\right)
+\frac{1}{U^2} \left(\frac{3}{4}|\gamma_0\rangle -\frac{1}{2}|m_{0;u}\rangle
-|m_{2;u}\rangle\right) \nonumber \\
&&+\frac{1}{U^3} \left(-\frac{9}{4}|\phi_{-1}\rangle 
-\frac{3}{4}|\gamma_1\rangle+\frac{25}{8}|m_{1;u}\rangle
+|m_{3;u}\rangle\right) \; , \\
|\Psi_2\rangle &=& |\gamma_1\rangle 
+\frac{1}{U} \left(\frac{1}{2}|\gamma_0\rangle -|m_{2;u}\rangle\right)
+\frac{1}{U^2} \left(-\frac{1}{4}|\phi_{-1}\rangle -\frac{3}{8}|\gamma_1\rangle
+\frac{1}{2}|m_{1;u}\rangle+|m_{3;u}\rangle\right)
\nonumber \\
&& +\frac{1}{U^3} \left(\frac{11}{4}|\gamma_0\rangle 
+\frac{1}{2}|\gamma_2\rangle-\frac{1}{2}|m_{0;u}\rangle
-\frac{7}{2}|m_{2;u}\rangle
-|m_{4;u}\rangle\right)\; .
\end{eqnarray}
Moreover, the diagonal and off-diagonal Lanczos parameters become
\begin{eqnarray}
e_0 &=& 
\frac{\langle\Psi_0 | \hat{\cal L} | \Psi_0\rangle}{\langle \Psi_0 | \Psi_0\rangle}
= -\frac{7}{4U^3} \nonumber \; ,\\
\tau_0 &=& 
-\sqrt{-\frac{\langle \Psi_1|\Psi_1\rangle}{\langle\Psi_0|\Psi_0\rangle}} 
= -\sqrt{1 + \frac{1}{4U^2}}=-\left(1+\frac{1}{8U^2}\right)
\; ,\\
e_1 &=& 
\frac{\langle \Psi_1 | \hat{\cal L} | \Psi_1\rangle}{\langle \Psi_1 | \Psi_1\rangle}
= -\frac{1}{2U} -\frac{31}{8U^3} \nonumber \; ,\\
\tau_1 &=& 
-\sqrt{-\frac{\langle \Psi_2|\Psi_2\rangle}{\langle\Psi_1|\Psi_1\rangle}} 
= -\sqrt{1 + \frac{3}{4U^2}}=-\left(1+\frac{3}{8U^2}\right)
\; .
\end{eqnarray}
For $n\geq 3$ the Lanczos matrix elements become constant,
and the Lanczos vectors obey a building principle.
This can be proven by induction.

\subsubsection{Induction formulae}

Up to and including third order in $1/U$ we have
\begin{eqnarray}
-\varepsilon_n=e_n= -\frac{1}{2U}-\frac{35}{8U^3}
&&\hbox{for $n\geq 2$}\; ,\nonumber \\
-t_n=\tau_n = -1-\frac{3}{8U^2} 
&& \hbox{for $n\geq 1$}\; ,
\end{eqnarray}
and, for $n\geq 3$,
\begin{eqnarray}
(-1)^n|\Psi_n\rangle &=& |\gamma_{n-1}\rangle 
+ \frac{1}{U} \left( \frac{1}{2} |\gamma_{n-2}\rangle - |m_{n;u}\rangle\right)
\nonumber \\
&& + \frac{1}{U^2} \left( -\frac{1}{4} |\gamma_{n-3}\rangle
+ a_n |\gamma_{n-1}\rangle
+\frac{1}{2} |m_{n-1;u}\rangle+|m_{n+1;u}\rangle\right)\\
&&+ \frac{1}{U^3} \left( \frac{1}{8} |\gamma_{n-4}\rangle 
+ b_n |\gamma_{n-2}\rangle
+\frac{1}{2} |\gamma_n\rangle -\frac{1}{2}|m_{n-2;u}\rangle
-c_n|m_{n;u}\rangle  -|m_{n+2;u}\rangle  \right)
\nonumber 
\end{eqnarray}
with $a_n=3(n-3)/8$, $b_n=3(n-4)/16+27/8$, and $c_n=3(n-4)/8+17/4$,
where we set $|\gamma_{-1}\rangle \equiv |\phi_{-1}\rangle$.

The induction proof is lengthy but straightforward and
can be found in Ref.~[\onlinecite{ruhl}].

\section{Hubbard bands in third order}
\label{sect:thirdorderbands}

In the last section we calculated the parameters for the 
hybridization function of the single-impurity Anderson
model in two-chain geometry up to and including third order in $1/U$.
The DMFT self-consistency equation on the Bethe lattice~(\ref{what-we-need})
shows that the hybridization function is identical to the 
impurity Green function which, in turn, is equivalent to the
local Green function of the Hubbard model.

\subsection{Density of states of the lower Hubbard band}

The hybridization function $\Delta(\omega)$ can be obtained from an 
equivalent scattering problem for a single particle on a semi-infinite chain.
We start this section by formulating this problem.
Next, we calculate the single-particle gap and 
the hybridization function in closed form. 
Lastly, we expand this expression systematically in $1/U$ 
which defines the `band-part' Green function.

\subsubsection{Scattering problem}

The Lanczos algorithm provides the
tridiagonal matrix representation of the hybridization function,
\begin{equation}
\mathfrak{h}_{\Delta}=
\left(
\begin{array}{cccccc}
		\varepsilon_{0} & t_{0} &  &  & &\\
		t_{0} & \varepsilon_{1} & \bar{t} & & & \\
		& \bar{t} & \bar{\varepsilon} & \bar{t} & & \\
	& &  \ddots & \ddots & \ddots & \\
	\end{array} 
\right)\; ,
\label{eq:matrixdarstellung} 
\end{equation}
where 
\begin{eqnarray}
\varepsilon_0=\frac{7}{4U^3} \quad , \quad
\varepsilon_1 &=& \frac{1}{2U}+\frac{31}{8U^3} \quad , \quad
\varepsilon_n =\frac{1}{2U}+\frac{35}{8U^3} \equiv \bar{\varepsilon} 
\quad (n\geq 2)\; , \nonumber \\
t_0 = 1 + \frac{1}{8U^2} \quad , \quad 
t_n &=& 1+\frac{3}{8U^2} \equiv \bar{t} 
\quad (n\geq 1)\; .
\end{eqnarray}
Note that only odd (even) orders appear in the $1/U$-expansion
of $\varepsilon_l$ ($t_l$).

As shown in Sect.~\ref{subsec:DMFT-selfconsistency},
the hybridization function can be obtained from
\begin{equation}
\Delta_{\rm LHB} (\omega) = 
\left(\left( \left(\omega + U/2 \right) \openone - 
\mathfrak{h}_{\Delta} - \imag \eta\right)^{-1} \right)_{00} \; .
\end{equation}
The matrix $\mathfrak{h}_{\Delta}$~(\ref{eq:matrixdarstellung}) 
corresponds to a tight-binding
Hamiltonian $\hat{K}$ which describes the transfer of a 
single particle on a semi-infinite chain,
compare eq.~(\ref{eq:perturbation}), plus 
a scattering potential $\hat{W}$ at the boundary of the chain,
\begin{eqnarray}
\hat{H}_{\rm scat}&=& \hat{K} +\hat{W}\nonumber \; ,\\
\hat{K}&=& \bar{t} \sum_{l=0}^{\infty} \left( |l\rangle\langle l+1|
+ |l+1\rangle\langle l|\right) 
+ \bar{\varepsilon} \sum_{l=0}^{\infty}|l\rangle\langle l|
\; , \label{defKandW}\\
\hat{W} &=& \varepsilon_0^* |0\rangle\langle 0| +
\varepsilon_1^* |1\rangle\langle 1| +
t_0^* \left( |0\rangle\langle 1| + |1\rangle\langle 0|\right) 
\nonumber 
\end{eqnarray}
with $\varepsilon_0^*=\varepsilon_0-\bar{\varepsilon}=-1/(2U)-21/(8U^3)$,
$\varepsilon_1^*=\varepsilon_1-\bar{\varepsilon}=-1/(2U^3)$,
$t_0^*=t_0-\bar{t}=-1/(4U^2)$.
Note that the scattering potential~$\hat{W}$ is attractive 
which results in a redshift of the density of states; see below.

The hybridization function can equally be calculated 
from the one-particle Hamiltonian $\hat{H}_{\rm scat}$
\begin{eqnarray}
\Delta_{\rm LHB} (\omega)&=& 
\left(\left( \left(\omega + U/2 \right) \openone - 
\mathfrak{h}_{\Delta} - \imag \eta\right)^{-1} \right)_{00}
\nonumber \\
&=& \frac{1}{2}\langle 0 | \left( \left(\omega + U/2 \right) \openone - 
\hat{H}_{\rm scat} - \imag \eta\right)^{-1} |0\rangle \; .
\end{eqnarray}
In this way, the Green function for the lower Hubbard band 
$G_{\rm LHB}(\omega)$ can be deduced from a one-particle problem.

\subsubsection{Single-particle gap}

The attractive $\hat{W}$ is too weak to generate a bound state
below the lower band edge of the tight-binding operator $\hat{K}$.
Therefore, it does not change the support of the imaginary part of
the bare Green function defined by $\hat{K}$
which is given by 
$|\omega-\bar{\epsilon}|\leq 2\bar{t}$.
In turn, this implies that the upper edge for the lower Hubbard
band is given by $\mu^-=-U/2+\bar{\epsilon}+2\bar{t}$ so that
the charge gap in~(\ref{eq:chargegap})
is given by (bandwidth $W=4$)
\begin{equation}
\Delta_{\rm c}= 2| -U/2+\bar{\epsilon}+2\bar{t}|
= U-4-\frac{1}{U}-\frac{3}{2U^2}-\frac{35}{4U^3} 
+{\cal O}\left(\frac{1}{U^4}\right) \; .
\end{equation}
The result to second order was derived earlier by Eastwood 
et alii.~\cite{kalinowski} 
Note that the coefficient to third
order is larger than anticipated in Ref.~[\onlinecite{kalinowski}]. 

Let $\Delta_{\rm c}(U_{\rm c})$ denote the critical 
value of the on-site interaction~$U$ where the 
charge gap closes, i.e., 
$\Delta_{\rm c}(U_{\rm c})=0$. Up to third order we find
\begin{eqnarray}
U_{\rm c}^{(0)}&=& 4\; , \nonumber \\
U_{\rm c}^{(1)}&=& 4.236 \; [5.90\%]\; , \nonumber \\
U_{\rm c}^{(2)}&=& 4.313 \; [1.82\%]\; , \nonumber \\
U_{\rm c}^{(3)}&=& 4.406 \; [2.16\%]\; , 
\end{eqnarray}
where the number in square brackets gives the percentage change
to the result of the previous order.
Apparently, the changes in the estimated critical interaction strength
from the second to the third order are of the same order of magnitude
and the critical interaction strength does not converge quickly
in these low orders.

\subsubsection{Green function for the scattering problem}

The calculation of the boundary Green function for a semi-infinite chain
with a local potential at the boundary is readily
accomplished.~\cite{Economou,Hoyer}
We define the general two-site Green functions
\begin{eqnarray}
g_{l,m}(\omega)&=& \langle l | 
\left(\left(\omega + U/2 \right) \openone - 
\hat{H}_{\rm scat} - \imag \eta\right)^{-1} |m \rangle \; , 
\label{fullGF}\\
g_{l,m}^{(0)}(\omega)&=& \langle l | 
\left(\left(\omega + U/2 \right) \openone - 
\hat{K} - \imag \eta\right)^{-1} |m \rangle \; .
\label{nonintGF}
\end{eqnarray}
The Green functions~(\ref{nonintGF}) 
for the tight-binding Hamiltonian~$\hat{K}$
are calculated explicitly in appendix~\ref{appF}.

In~(\ref{fullGF}) we use the operator identify $(\hat{A}-\hat{B})^{-1}=
\hat{A}^{-1} + \hat{A}^{-1}\hat{B}(\hat{A}-\hat{B})^{-1}$
with $\hat{A}=(\omega+U/2)\openone-\hat{K}$ and $\hat{B}=\hat{W}$
so that we can write
\begin{eqnarray}
g_{0,0}(\omega) \!&=&\! g_{0,0}^{(0)}(\omega) 
+\sum_{l,m=0}^{\infty} g_{0,l}^{(0)}(\omega) \langle l|\hat{W} |m\rangle
g_{m,0}(\omega) 
 \label{firstoftwo}
\\
\!&=&\! g_{0,0}^{(0)}(\omega) + 
\epsilon_0^* g_{0,0}^{(0)}(\omega) g_{0,0}(\omega) 
+
\epsilon_1^* g_{0,1}^{(0)}(\omega) g_{1,0}(\omega) 
+
t_0^* \left(g_{0,0}^{(0)}(\omega) g_{1,0}(\omega) 
+ g_{0,1}^{(0)}(\omega) g_{0,0}(\omega) \right)
\, ,\nonumber 
\end{eqnarray}
where we used the locality of the scattering potential 
$\hat{W}$~(\ref{defKandW}) in the second step.
Likewise we obtain 
\begin{equation}
g_{1,0}(\omega) =
g_{1,0}^{(0)}(\omega) + 
\epsilon_0^* g_{1,0}^{(0)}(\omega) g_{0,0}(\omega) 
+
\epsilon_1^* g_{1,1}^{(0)}(\omega) g_{1,0}(\omega) 
+
t_0^* \left(g_{1,1}^{(0)}(\omega) g_{0,0}(\omega) 
+ g_{1,0}^{(0)}(\omega) g_{1,0}(\omega) \right)
\; . \label{secondoftwo} 
\end{equation}
We can solve the coupled equations~(\ref{firstoftwo}) and (\ref{secondoftwo})
to give our final result ($g_{0,0}(\omega)=2G_{\rm LHB}(\omega)$)
\begin{eqnarray}
g_{0,0}(\omega)&=& 
\frac{g_{0,0}^{(0)}(\omega) -\varepsilon_1^*
\left[
g_{0,0}^{(0)}(\omega) g_{1,1}^{(0)}(\omega) 
- g_{0,1}^{(0)}(\omega) g_{1,0}^{(0)}(\omega)
\right]}{N(\omega)}
\nonumber \; ,\\
N(\omega) &=& 
1-\varepsilon_0^* g_{0,0}^{(0)}(\omega) 
-t_0^*(g_{1,0}^{(0)}(\omega) +g_{0,1}^{(0)}(\omega) )
-\varepsilon_1^* g_{1,1}^{(0)}(\omega) 
\label{eq:finalclumsy}\\
&& +
\left(\varepsilon_0^*\varepsilon_1^*-(t_0^*)^2\right)
\left(g_{0,0}^{(0)}(\omega)g_{1,1}^{(0)}(\omega)
        -g_{1,0}^{(0)}(\omega)g_{0,1}^{(0)}(\omega) \right)
\; .
\nonumber
\end{eqnarray}
{}From appendix~\ref{appF} it follows that
$g_{1,0}^{(0)}(\omega)=g_{0,1}^{(0)}(\omega)$
and
\begin{eqnarray}
g_{1,1}^{(0)}(\omega)&=& g_{0,0}^{(0)}(\omega) 
+ \bar{t}^2 \left(g_{0,0}^{(0)}(\omega)\right)^3 \nonumber \; ,\\
\bar{t}^2 \left(g_{0,0}^{(0)}(\omega)g_{1,1}^{(0)}(\omega)
        -g_{1,0}^{(0)}(\omega)g_{0,1}^{(0)}(\omega)\right)
&=& \bar{t}g_{1,0}^{(0)}(\omega)=\left[\bar{t}g_{0,0}^{(0)}(\omega)\right]^2
\label{eq:gfzauber}
\end{eqnarray}
so that we can cast our final third-order result into the form
\begin{equation}
g_{0,0}(\omega)= 
\frac{
g_{0,0}^{(0)}(\omega) -\varepsilon_1^*\left[g_{0,0}^{(0)}(\omega)\right]^2}%
{1
-\left(\varepsilon_0^* +\varepsilon_1^*\right)g_{0,0}^{(0)}(\omega) 
+\left(\varepsilon_0^*\varepsilon_1^*-(t_0^*)^2-2t_0^*\bar{t}\right)
\left[g_{0,0}^{(0)}(\omega)\right]^2
-\varepsilon_1^* \bar{t}^2 \left[g_{0,0}^{(0)}(\omega)\right]^3
}
\; .
\label{eq:final}
\end{equation}
The density of states is the imaginary part of this expression,
$2\pi D_{\rm LHB}(\omega)={\rm Im}[g_{00}(\omega)]$.
The bare boundary Green function is given by 
[$x=(\omega+U/2-\bar{\varepsilon})/(2\bar{t})$]
\begin{equation}
\bar{t}g_{0,0}^{(0)}(\omega) = \Theta\left(x^2-1\right)
\left(x-\sgn(x)\sqrt{x^2-1}\right) 
+ \Theta\left(1-x^2\right) \left[ x +\imag \sqrt{1-x^2}\right]\; ,
\end{equation}
where $\Theta(x)$ is the Heaviside step-function.
For the density of states we only need the region $|x|\leq 1$.

\begin{figure}[ht]
\includegraphics{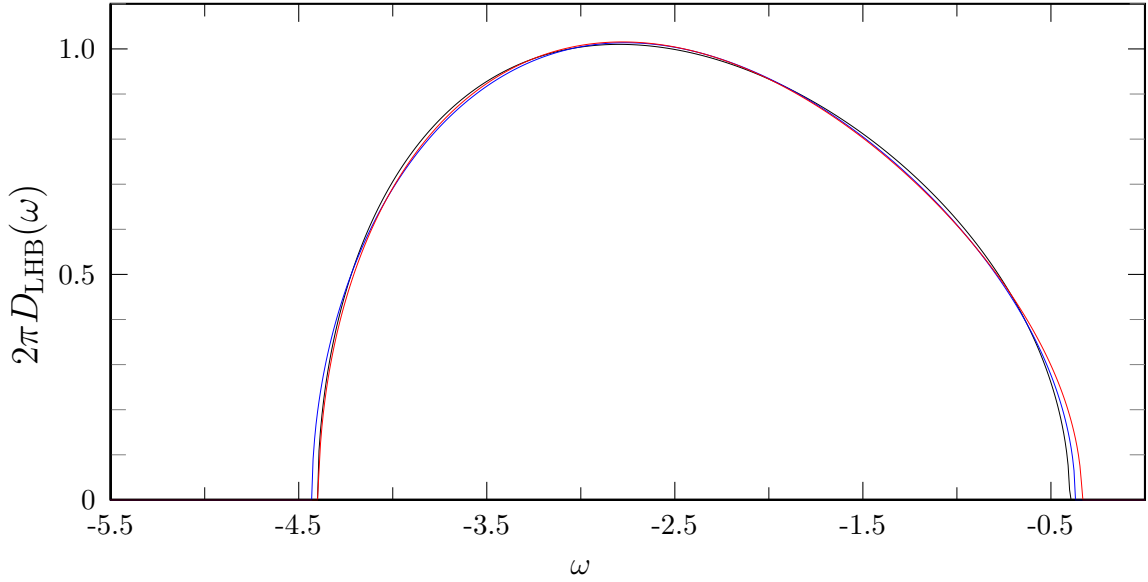}
	\caption{Density of states of the lower Hubbard band, 
$\pi D_{\rm LHB}^{[n]}(\omega)$ for $U=5$ (bandwidth $W=4$)
up to and including orders $n=1,2,3$ (black, blue, red colors).
	\label{fig:densityofstatesdependenceonu}}
\end{figure}

In Fig.~\ref{fig:densityofstatesdependenceonu} we show
the results for the density of states of the lower Hubbard band
for $U=5$ (bandwidth $W=4$)
to first, second, and third order in $1/U$.
The overall spectra display a redshift 
of the Hubbard semi-ellipse~(\ref{eq:semiellipse}) which 
describes the density of states to leading-order,
$2D_{\rm LHB}^{(0)}(\omega)=\rho(\omega+U/2)$.
The spectra to higher orders differ from each other mostly
by a shift in the spectral support so that the deviations
are best visible close to the band edges.

\subsubsection{Band part of the Green function}

The full solution~(\ref{eq:finalclumsy}) contains higher-order corrections
in $1/U$ due to the interaction-dependence of the denominator~$N(\omega)$.
We may expand it order-by-order to derive a Taylor series in~$1/U$
for the Green function. Such an order-by-order expansion ignores the fact that
the attractive potential~$\hat{W}$ generates resonance-contributions
at the band edges of the Hubbard band; see below.
Therefore, we denote the Green function from
the order-by-order expansion as `band-part' Green function.
It can be cast into the form
\begin{eqnarray}
2\bar{t} G_{\rm LHB}^{\rm band}(\omega) &=& \sum_{n=0}^3 \lambda_n 
(\widetilde{g}_n(x)+g_n(x)) \;, \nonumber \\
2\left(1+\frac{3}{8U^2}\right)x 
&=& \omega+\frac{U}{2} -\frac{1}{2U}-\frac{35}{8U^3}\; ,\nonumber \\
\widetilde{g}_n(x) &=& \Theta\left(x^2-1\right)
\left(T_{n+1}(x)-\sgn(x)\sqrt{x^2-1}U_n(x)\right) \; , \nonumber\\
g_n(x) &=& \Theta\left(1-x^2\right)
\left(T_{n+1}(x)+\imag \sqrt{1-x^2}U_n(x)\right) \; ,
\label{eq:bandpart}
\end{eqnarray}
with $T_n(x)$ [$U_n(x)$] as the Chebyshev polynomials
of the first [second] kind~\cite{Abramovitz} and 
\begin{eqnarray}
\lambda_0= 1 &,& \lambda_1= -\frac{1}{2U}-\frac{39}{16U^3}\; , \nonumber \\
\lambda_2= -\frac{1}{4U^2} &,& \lambda_3= -\frac{1}{8U^3}
\end{eqnarray}
are the expansion coefficients.
The first-order result was derived earlier in Ref.~[\onlinecite{Nahrgang}].
Using an intuitive method,
Eastwood et al.~\cite{kalinowski} 
derived the `band-part' Green function to second order in $1/U$
for the Hubbard model in infinite dimensions.
So far, their method could not be extended systematically to higher orders.

\begin{figure}[ht]
\includegraphics[width=0.95 \textwidth]{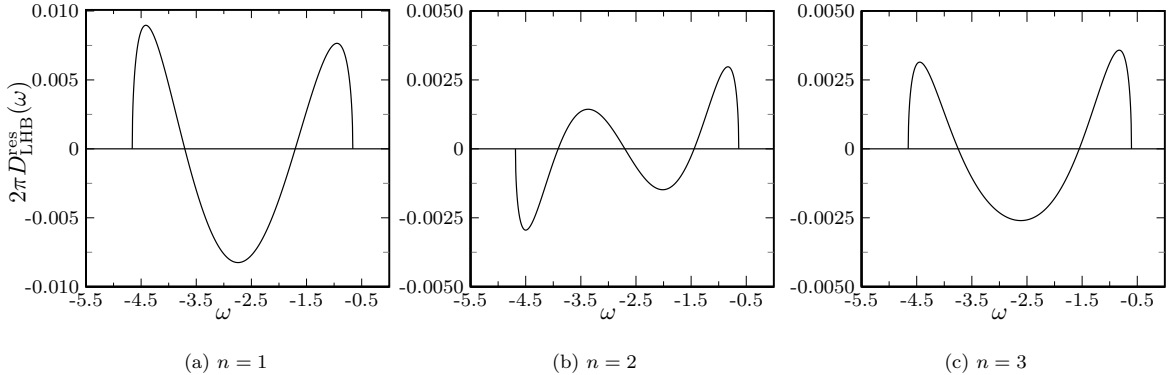}
	\caption{Resonance contribution to the density of states 
of the lower Hubbard band, $D_{\rm LHB}^{\rm res}(\omega)=
D_{\rm LHB}(\omega)-D_{\rm LHB}^{\rm band}(\omega)$,
as a function of frequency in $n$th order perturbation theory 
for $U=5.5$ (bandwidth $W=4$).
\label{fig:densityofstates-bandpart}}
\end{figure}

In Fig.~\ref{fig:densityofstates-bandpart}
we show the resonance contribution to the density of states,
$D_{{\rm LHB}}^{\rm res}(\omega)$ for $U=5.5$.
It is defined as 
the difference between the band part~(\ref{eq:bandpart}),
$D_{\rm LHB}^{\rm band}(\omega)$, and the full 
density of states $D_{\rm LHB}(\omega)$~(\ref{eq:final}).
The difference is seen to be fairly small which had to
be expected because the potential~$\hat{W}$ is rather weak.
In general, the resonance contributions
increase slightly the density of states close to
the band edges and decrease it in the middle of the band.

\subsection{Comparison with numerical results}

Finally, we compare our analytical results with data of advanced
numerical methods for the DMFT for the Mott--Hubbard insulator.
The Dynamical Density-Matrix Renormalization Group (DDMRG)
method provides the gap and the density of states
at zero temperature.~\cite{Nishimoto}  
Quantum Monte-Carlo (QMC) gives the
Matsubara Green function at low but finite temperatures.
Ref.~[\onlinecite{kalinowski}] contains a comparison
with early methods in the field.

\begin{figure}[ht]
\includegraphics[width=0.9\textwidth]{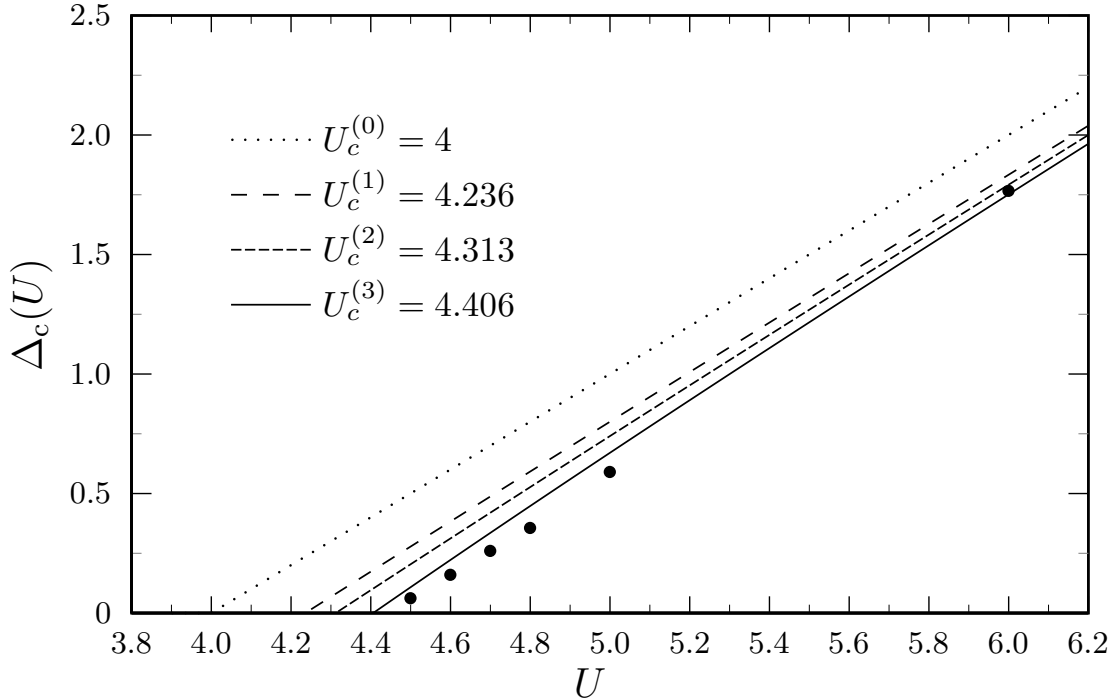}
	\caption{Charge gap as a function of the interaction strength
for various orders in the $1/U$-expansion.
The dots are DDMRG data points.~\cite{Nishimoto}
\label{fig:gap-ddmrg}}
\end{figure}

\subsubsection{Gap}

In Fig.~\ref{fig:gap-ddmrg} we show the gap as a function of the 
interaction strength for various orders in the $1/U$-expansion 
together with the DDMRG data of Ref.~[\onlinecite{Nishimoto}].
The third-order theory reproduces the DDMRG data points very well.

Note, however, that in another DDMRG study~\cite{Uhrig} the gap closes
around $U=4.8$. The differences in the two approaches lies in the
reconstruction of the density of states and the
extrapolation of the gap from the finite-size data.
Apparently, different reconstruction algorithms
can result in substantially different extrapolations
close to the transition. 

\begin{figure}[ht]
\includegraphics[width=0.95 \textwidth]{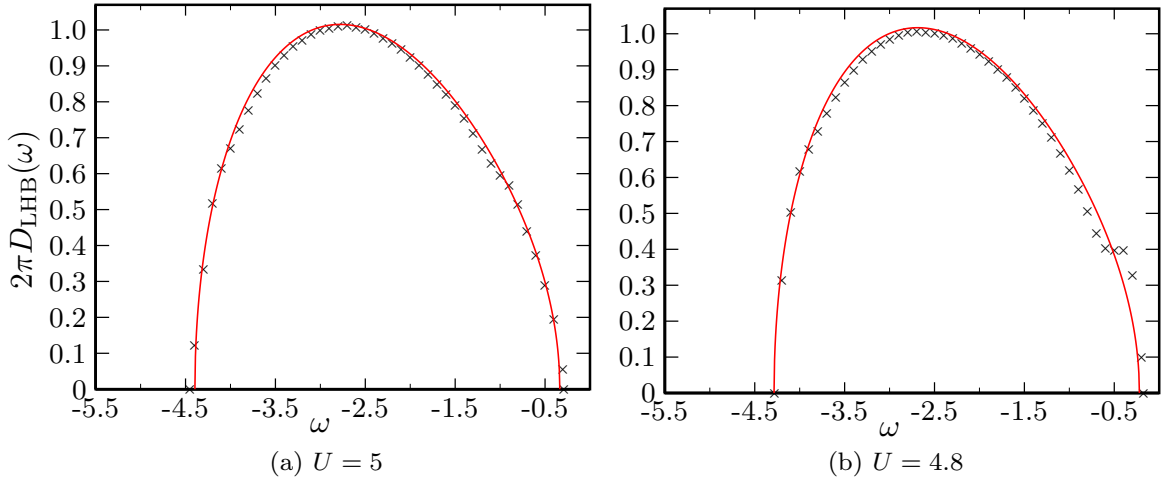}
	\caption{Density of states of the lower Hubbard band
from third-order perturbation theory in $1/U$ (full line)
in comparison with DDMRG data points~\cite{Nishimoto}
for $U=5$ and $U=4.8$.\label{fig:dos-ddmrg}}
\end{figure}

\subsubsection{Lower Hubbard band}

In Fig.~\ref{fig:dos-ddmrg} we show the density of states 
for $U=4.8$ to third order in $1/U$
together with the DDMRG data of Ref.~[\onlinecite{Nishimoto}].
The overall agreement is very good.
This has already been observed 
from the results to second order.~\cite{kalinowski}

It is seen, though, that
a resonance develops at the upper band edge in the DDMRG data
which is not seen in perturbation theory to third order.
For $U=4.5$, the resonance is more pronounced~\cite{Nishimoto}
and resembles the split quasi-particle peak of the metallic phase.
One may wonder whether such a resonance could be obtained from
higher-order perturbation theory. A model study~\cite{Hoyer}
shows that the parameter set
$\varepsilon_0^*=-0.2$ and $\varepsilon_{1\leq m\leq 9}^*=0.1/m$
in the scattering potential $\hat{W}$ can readily account
for both the overall redshift of the density of states
and a resonance at the upper band edge. Since the range
of the repulsive potential is finite, an expansion
of the density of states to high but finite order
could possibly reproduce the resonance seen in the DDMRG data.

\begin{figure}[ht]
\includegraphics{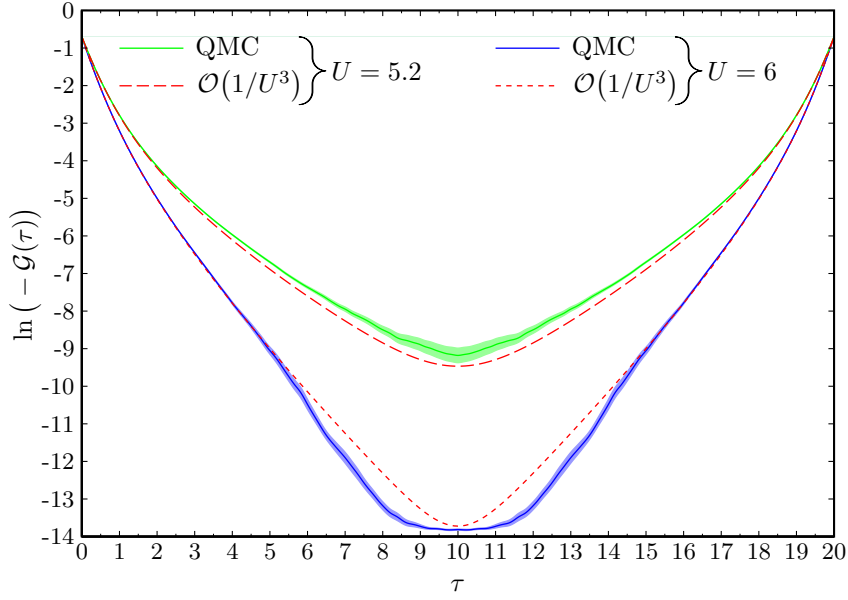}
\caption{Third-order result and QMC data
for the Matsubara Green function 
for $U=6$ (blue) and $U=5.2$ (green). The inverse temperature is $\beta=20$
($T=0.05$), the gaps are $\Delta_{\rm c}(U=6)=1.75$
and $\Delta_{\rm c}(U=5.2)=0.89$, respectively.
Note that the data are shown on a logarithmic scale.
The shading indicates the statistical error in the QMC data.
\label{fig:qmc-nils-alles}}
\end{figure}

\subsubsection{Matsubara Green function}

The Matsubara Green function for the Hubbard model is defined by
\begin{equation}
{\cal G}(\tau)=-\frac{1}{L} \sum_i  {\rm Tr}\left[
e^{\beta(\Omega-\hat{H})}{\cal T}_{\tau} 
\hat{c}_{i,\sigma}(\tau)\hat{c}_{i,\sigma}^+(0)\right] \; ,
\end{equation}
where $\beta=1/k_{\rm B}T$ is the inverse temperature,
$\Omega$ is the grand-canonical potential, and ${\cal T}_{\tau}$
orders the operators in imaginary time. The operators
in imaginary-time Heisenberg representation are defined 
by ($-\beta \leq \tau \leq \beta$)
\begin{equation}
\hat{c}_{i,\sigma}(\tau)= e^{\tau \hat{H}} \hat{c}_{i,\sigma}e^{-\tau \hat{H}}
\quad ,\quad
\hat{c}_{i,\sigma}^+(\tau)= e^{\tau \hat{H}} \hat{c}_{i,\sigma}^+e^{-\tau \hat{H}}
\; .
\end{equation}
The Fourier transformation of the Matsubara Green function 
is defined on the points $\imag\omega_n=(2n+1)\pi/\beta$ ($n$: integer)
on the imaginary axis. 

The retarded Green function at finite temperature~$T$ is obtained
from the analytic continuation,
\begin{equation}
G_{\rm ret}(\omega;\beta) = {\cal G}(\imag\omega_n \to \omega 
+ \imag \eta)\; .
\end{equation}
Therefore, we may express the Matsubara Green function
with the help of the density of states at finite temperature
in the form
\begin{equation}
{\cal G}(\tau)= 
\int_{-\infty}^{\infty} {\rm d}\omega
\left[\frac{{\rm Im}\left[G_{\rm ret}(\omega;\beta)\right]}{\pi}\right]
\frac{e^{-\omega\tau}}{e^{-\beta\omega}+1}
\label{eq:easyway}
\end{equation}
with $0\leq \tau\leq \beta$.
Note that it is easy to evaluate~(\ref{eq:easyway})
for a given density of states but it is very difficult
to reconstruct the density of states from numerical data
for ${\cal G}(\tau)$.

\begin{figure}[ht]
\includegraphics[height=6cm]{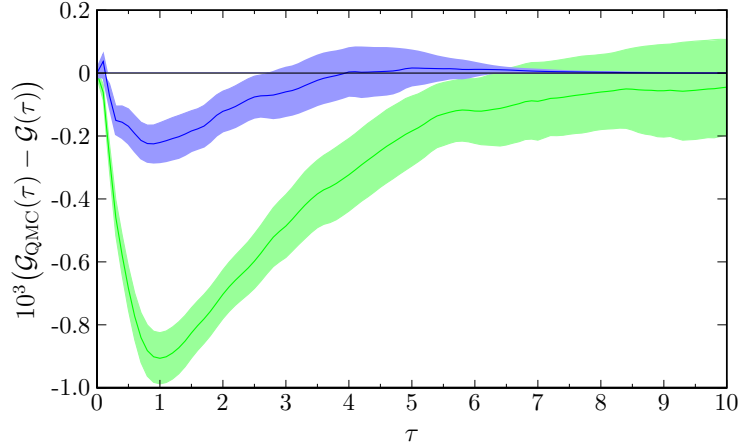}
\caption{Difference between the third-order result and QMC data 
for the Matsubara Green function
for $U=6$ (blue) and $U=5.2$ (green). The inverse temperature is $\beta=20$
($T=0.05$), the gaps are $\Delta_{\rm c}(U=6)=1.75$
and $\Delta_{\rm c}(U=5.2)=0.89$, respectively.
Note that the difference is augmented by a factor of $10^3$
to make it visible. The shading indicates the statistical error 
in the QMC data.
\label{fig:qmc-nils}}
\end{figure}

For very low temperatures and for large interaction strengths
we approximate
the density of states by its zero-temperature expression 
to third order,
\begin{equation}
{\cal G}(\tau) = - \int_{-\infty}^{\infty} {\rm d}\omega
\Bigl[ D_{\rm LHB}(\omega)+D_{\rm LHB}(-\omega)\Bigr]
\frac{e^{-\omega\tau}}{e^{-\beta\omega}+1}
\label{eq:comparenils}
\end{equation}
which we compare with QMC data of N.\ Bl\"umer.~\cite{nils}
The approximation is not as drastic as it may seem because,
deep in the Mott--Hubbard insulator, thermal excitations
are exponentially suppressed due to the finite charge gap.
Therefore, corrections to~(\ref{eq:comparenils}) should be exponentially
small in $\Delta_{\rm c}(U)/k_{\rm B}T$.

In Figs.~\ref{fig:qmc-nils-alles} and~\ref{fig:qmc-nils}
we compare our analytical results~(\ref{eq:comparenils})
to QMC data for $\beta=20$ ($T=0.05$) at $U=6$ and $U=5.2$
where the gaps are $\Delta_{\rm c}(U=6)=1.75$
and $\Delta_{\rm c}(U=5.2)=0.89$, respectively.
The results agree very well. Note, however, that
${\cal G}(\tau)$ is rather feature-less so that
fine points such as the width of the Hubbard bands or
the density of states cannot be reconstructed easily from QMC data
for ${\cal G}(\tau)$.

\section{Conclusions}
\label{sec:conclusions}

In this work, we have studied the Mott--Hubbard insulating phase
of the Hubbard model on a Bethe lattice
with infinite coordination number. 
We have adopted the Kato--Takahashi perturbation theory to solve the 
self-consistency equation of the Dynamical Mean-Field Theory
for the symmetric single-impurity Anderson model
in perturbation theory up to and including third order in the inverse
coupling strength~$U$. To this end it has been necessary to
use the mapping of the single-impurity Anderson model
from the `star geometry' onto the `two-chain geometry'
which represents the energetically separated lower and upper Hubbard
bands. In higher orders, a multi-chain mapping is required in order
to resolve the various Hubbard sub-bands. For the present study,
we could ignore the secondary Hubbard bands whose weight is
of fourth order in $1/U$.

Our results for the Mott--Hubbard gap reproduce 
those of an earlier analytic study~\cite{kalinowski}  
of the Hubbard model 
on the Bethe lattice with infinite coordination number. We confirm
the second-order results of Ref.~[\onlinecite{kalinowski}] and extend them
to third order systematically.
The agreement between the perturbation theory
in~$1/U$ and the DDMRG data of Ref.~[\onlinecite{Nishimoto}] for the gap
is very good.
Note, however, that the precise value of the critical interaction~$U_{\rm c}$
where the gap closes, 
and the analytical behavior of the gap as a function of~$U$ 
close to the transition are still under debate.~\cite{Uhrig}

The previous study~\cite{kalinowski}
provides the Green function as a Taylor expansion in~$1/U$
whereas the present study includes resonance corrections.
The full density of states results from
the calculation of the boundary Green function 
for a particle on a semi-infinite chain with nearest-neighbor electron
transfers and an attractive interaction at and near the boundary,
whose parameters we derived to third order in $1/U$.
For all interaction strengths where perturbation theory is
applicable, $U\gtrsim 5$ (bandwidth: $W=4$, $4.4 \lesssim U_{\rm c}
\lesssim 4.8$), the resonance contributions are small.

For $U\gtrsim 5$,
the agreement between the analytical results
for the density of states and the DDMRG data~\cite{Nishimoto}
is very good for all frequencies.
In addition, our zero-temperature expressions for the density of states
provides a very good approximation for the density of states
at small but finite temperatures. This can be seen from the
excellent agreement between our approximate Matsubara Green function
and Quantum Monte-Carlo data.~\cite{nils}

As in all kinds of perturbation theories, the number of terms to be
calculated rapidly increases with the index of the order. 
In principle, the fourth-order terms could
still be calculated `by hand'. This requires a four-chain geometry
so that the secondary Hubbard sub-bands can be treated properly.
To fourth order 
there are more than 30~terms in the Kato--Takahashi operator
and in the projected Hamiltonian.
According to our analysis, much higher orders are needed to
reproduce a resonance feature seen in the DDMRG data~\cite{Nishimoto}
at the upper band edge of the lower Hubbard band.
Such high-order calculations for the density of states 
appear to be forbiddingly costly within the DMFT.

The ground-state energy of the Hubbard model 
on the Bethe lattice with infinite coordination number was calculated
to high orders using a computer algorithm based on the
Kato--Takahashi expansion.~\cite{eva}
In the future, we plan to devise a similar algorithm 
for the calculation of the Mott--Hubbard gap.
With a high-order expansion for the Mott--Hubbard gap we should be able
to locate $U_{\rm c}$ with a much better accuracy.

\begin{center}
{\bf Acknowledgments}
\end{center}
We thank Marlene Nahrgang for her contributions
to the early stages of this work,
and J\"org B\"unemann for useful discussions.

\appendix

\section{Weight of the secondary lower Hubbard band}
\label{appD}

We apply particle-hole symmetry and the self-consistency
equation~(\ref{what-we-need}) to
the sum rule for the density of states~\cite{FetterWalecka} and find
\begin{equation}
\frac{1}{2} = \frac{1}{\pi} \int_{-\infty}^0 {\rm d}\omega
{\rm Im}\left[\Delta(\omega)\right]
= \sum_{i=0}^{\infty}\biggl[\sum_m V_{i,m}^2 \biggr]
= \sum_{i=0}^{\infty}g_i \; ,
\label{eq:sumrule}
\end{equation}
where we use eq.~(\ref{eq:delta-subbands}) in the second step.
The $i$th sub-band of the Hubbard band contributes 
the weight $g_i$.

The weights $g_i$ can be calculated perturbatively. From the definition
of the Green function of the lower Hubbard band~(\ref{eq:gfsiamtransformed1})
we can readily write the sum rule~(\ref{eq:sumrule})
as
\begin{equation}
\frac{1}{2} = \frac{1}{\pi} \int_{-\infty}^0 {\rm d}\omega
{\rm Im}\left[G_{\rm LHB}(\omega)\right]
= \langle \Phi | 
\hat{\Gamma}_{0,L}^+ \hat{d}_{\uparrow}^+ 
\hat{d}_{\uparrow} \hat{\Gamma}_{0,L}| \Phi\rangle \; .
\end{equation}
The state $|\overline{\Psi}\rangle= 
\hat{d}_{\uparrow} \hat{\Gamma}_{0,L}| \Phi\rangle$
can readily be calculated from the series expansion
of the operator $\hat{\Gamma}_{0,L}$, eq.~(\ref{gammatothirdorder}),
applied to the state~$|\Phi\rangle$, see
eqs.~(\ref{eq:lanczosstartingvector})-(\ref{eq:groundstatedownhnaught}).
Up to and including third order in $1/U$, we find
\begin{equation}
\hat{P}_{0,L-1}^{(0)} |\overline{\Psi}\rangle =
\left(1-\frac{1}{2U^2}\right) |\phi_{-1}\rangle 
-\frac{1}{U} \left(1+\frac{1}{U^2}\right) |m_{0;u}\rangle
+\frac{1}{U^2}|m_{1;u}\rangle
-\frac{1}{U^3}|m_{2;u}\rangle
\label{eq:langnet}
\end{equation}
in the subspace of the ground-states of $\hat{H}_0$ for $L-1$ particles
on $L$ lattice sites.
In addition, there is a finite second-order contribution in the 
subspace with excitation energy~$U$ above the ground states of $\hat{H}_0$,
\begin{equation}
\hat{P}_{1,L-1}^{(0)} 
|\overline{\Psi}\rangle =
\frac{1}{U^2} \left( \frac{1}{2}|\phi_{0;u}^*\rangle 
+ \frac{1}{2}|\phi_{0;d}^*\rangle + |\chi_{0}^*\rangle 
\right) +{\cal O}\left(\frac{1}{U^3}\right) \; ,
\end{equation}
where
\begin{eqnarray}
|\phi_{0;u}^*\rangle &=& 
\sqrt{\frac{1}{2}} 
\hat{\beta}_{0,\downarrow}^+ \hat{\alpha}_{0,\downarrow}
\prod_{l=0}^{(L-3)/2} 
\hat{\alpha}_{l,\uparrow}^+\hat{\alpha}_{l,\downarrow}^+
|\hbox{vac}\rangle
\;, \nonumber \\
|\phi_{0;d}^*\rangle &=& 
-\sqrt{\frac{1}{2}} \hat{\beta}_{0,\uparrow}^+ \hat{\alpha}_{0,\uparrow}
\prod_{l=0}^{(L-3)/2} 
\hat{\alpha}_{l,\uparrow}^+\hat{\alpha}_{l,\downarrow}^+
|\hbox{vac}\rangle
\; ,\\
 |\chi_{0}^*\rangle &=& 
-\sqrt{\frac{1}{2}} \hat{\beta}_{0,\downarrow}^+ \hat{\alpha}_{0,\uparrow}
\prod_{l=0}^{(L-3)/2} 
\hat{\alpha}_{l,\uparrow}^+\hat{\alpha}_{l,\downarrow}^+
|\hbox{vac}\rangle
\;. \nonumber 
\end{eqnarray}
Therefore, the weight of the secondary sub-band, $i=1$, is given by
\begin{equation}
g_1=\langle \overline{\Psi} |\hat{P}_{1,L-1}^{(0)} |\overline{\Psi} \rangle
=\frac{1}{2U^4} \left(\frac{1}{4} + \frac{1}{4}+ 1\right) =\frac{3}{4U^4} \; ,
\label{eq:needprimarysubbandonly}
\end{equation}
and corrections are of higher order in $1/U$.
Because the higher sub-bands are even smaller in weight,
$g_{i\geq 2}={\cal O}(U^{-5})$, we can conclude from the sum-rule
that $g_0=1/2-3/(4U^4)$. The direct calculation
of $g_0$ from eq.~(\ref{eq:langnet}) is not possible because
it lacks the fourth-order correction $-7/(4U^4)|\phi_{-1}\rangle$.

Equation~(\ref{eq:needprimarysubbandonly}) shows that,
up to and including third order in $1/U$,
only the primary Hubbard sub-band contributes to the density of states
for $\omega<0$.

\section{Green functions for a semi-infinite tight-binding chain}
\label{appF}

Let $\hat{B}=\sum_{l=0}^{\infty}|l\rangle\langle l+1|
+|l+1\rangle\langle l|$ describe the motion of a single particle
over a semi-infinite chain.
By definition, we have $\hat{B}|0\rangle =|1\rangle$ and
$\hat{B}|l\rangle =|l+1\rangle + |l-1\rangle$ for $l\geq 1$.
By induction we can prove the following lemma:
\begin{itemize}
\item[(i)]
For all  $n \in \mathbb{N}_0$ we find 
\begin{equation}
U_n(\hat{B}/2) |0\rangle = |n\rangle\; .
\end{equation}
Here, $U(x)$ are the Chebyshev polynomials of the second 
kind.~\cite{Abramovitz}
\item[(ii)]
For all $n,l \in \mathbb{N}$ we have
\begin{equation}
U_n(\hat{B}/2) |l\rangle 
= \sum_{i=0}^{n} \Theta_{2(l-n)+2i} |l-n+2i\rangle \; .
\end{equation}
Here, $\Theta_l$ denotes the discrete unit-step function 
$\Theta: \mathbb{Z} \rightarrow \{0,1\}$,
\begin{equation}\label{eq:discreteunitstep}
\Theta_l := \left\{ \begin{array}{rll}
1 & \hbox{for} & l\geq 0\; , \\ 
0 & \hbox{for} & l<0\; .
\end{array}
\right.
\end{equation}
\end{itemize}
For the imaginary part of the bare Green function~(\ref{nonintGF})
we can write ($\widetilde{\omega}=\omega+U/2-\bar{\varepsilon}$)
\begin{equation}
{\rm Im}\left[g_{l,m}^{(0)}(\omega)\right]= 
\pi \langle l | \delta\left(
\widetilde{\omega}\openone -\bar{t}\hat{B} \right) |m \rangle \; ,
\end{equation}
which is finite in the interval $|\widetilde{\omega}|\leq 2\bar{t}$.
We set $x=\widetilde{\omega}/(2\bar{t})$ and obtain
\begin{equation}
{\rm Im}\left[g_{l,m}^{(0)}(\omega)\right]= 
\frac{\pi}{2\bar{t}} \langle l | \delta\left(
x\openone -\hat{B}/2 \right) |m \rangle \; .
\end{equation}
When we formally expand the `function' $f(x)=\pi\delta(x-z)$ 
($|x|\leq 1$, $|z|\leq 1$) in a Chebyshev
series,~\cite{ruhl}
\begin{equation}
\pi\delta(x-z) =2\sqrt{1-x^2}\sum_{n=0}^{\infty}U_n(x)U_n(z) \; ,
\end{equation}
we can write the imaginary part of the bare Green function
as
\begin{equation}
{\rm Im}\left[g_{l,m}^{(0)}(\omega)\right]= 
\frac{1}{\bar{t}}\Theta\left(1 -x^2\right) 
\sqrt{1-x^2}\sum_{n=0}^{\infty}U_n(x)
\langle l | U_n(\hat{B}/2) |m \rangle \; .
\end{equation}
Because $\hat{B}$ is Hermitian so that
$g_{l,m}^{(0)}(\omega)= g_{m,l}^{(0)}(\omega)$, we can restrict ourselves
to $l=m+h$ with $h\in \mathbb{N}_0$.
With the help of the lemma, it is not difficult to show that
for $h,m\in \mathbb{N}_0$,
\begin{eqnarray}
\label{eq:imagpartk} 
\bar{t} {\rm Im}\left[ g_{m+h,m}^{(0)}(\omega)\right]
&=& \Theta(1-x^2) \sqrt{1-x^2}
\Bigl\{ \delta_{m,0} U_h(x) \nonumber \\ 
&& + (1-\delta_{m,0}) \sum_{k,y=0}^{\infty} 
\Theta_{m+y-k} \Theta_{k-y} 
\left[ U_{2k}(x)\delta_{h,2y} + U_{2k+1}(x)\delta_{h,2y+1} \right]
\Bigr\} 
\end{eqnarray}
with the abbreviation $x=(\omega+U/2-\bar{\varepsilon})/(2\bar{t})$, 
the discrete unit-step function $\Theta_l$, see~(\ref{eq:discreteunitstep}),
and $\Theta(x)$ as the Heaviside step-function.

The real part follows from the Kramers--Kronig 
transformation.~\cite{FetterWalecka}
We find~\cite{ruhl}
\begin{equation}
\label{eq:realpartk} 
\bar{t} {\rm Re}\left[ g_{m+h,m}^{(0)}(\omega)\right]
= \delta_{m,0} I_h(x)
+ (1-\delta_{m,0}) \sum_{k,y=0}^{\infty} 
\Theta_{m+y-k} \Theta_{k-y} 
\left[ I_{2k}(x)\delta_{h,2y} + I_{2k+1}(x)\delta_{h,2y+1} \right]
\; .
\end{equation}
As is proven by induction in Ref.~[\onlinecite{ruhl}],
we have ($n\in \mathbb{N}_0, p\in \mathbb{N}$),
\begin{eqnarray}
I_n(x) &=& T_{n+1}(x) \quad \hbox{for} \quad  |x|\leq 1\; , \nonumber \\
I_n(x) &=& T_{n+1}(x) -\sgn(x)\sqrt{x^2-1} U_n(x) 
\quad \hbox{for} \quad  |x|\geq 1\;,\\
{}[I_n(x)]^p &=& T_{p(n+1)}(x) -\sgn(x)\sqrt{x^2-1} U_{p(n+1)-1}(x) 
\quad \hbox{for} \quad  |x|\geq 1\;,\nonumber 
\end{eqnarray}
where $T_n(x)$ are the Chebyshev polynomials of the first 
kind.~\cite{Abramovitz}
In particular, the bare boundary Green function reads ($\omega< 0$)
\begin{equation}
\bar{t}g_{0,0}^{(0)}(\omega) = \Theta\left(x^2-1\right)
\left(x-\sgn(x)\sqrt{x^2-1}\right) 
+ \Theta\left(1-x^2\right) \left[ x + \imag \sqrt{1-x^2}\right]\; .
\end{equation}
Finally, we note that powers of the bare Green function obey
for $n \in \mathbb{N}_0, p \in \mathbb{N}$
\begin{equation}
\bar{t}^p g_{n,0}^{p}(\omega) = \Theta(x^2-1) I_{n}^p(x) 
+ \Theta(1-x^2) \left\{ T_{p(n+1)}(x) + \imag \sqrt{1-x^2}
U_{p(n+1)-1}(x) \right\}\; .\label{eq:productgfscattering} 
\end{equation}
With these relations and the properties of the Chebyshev polynomials,
one can readily prove eq.~(\ref{eq:gfzauber}); for details, see
Ref.~[\onlinecite{ruhl}].

\end{document}